\documentclass[aps,10pt,twocolumn,english,showpacs,floatfix,amsmath,amssymb,prb,superscriptaddress]{revtex4-1}
\usepackage{graphicx}
\usepackage{graphics}
\usepackage{amsmath}
\usepackage{amsfonts}
\usepackage{amssymb}
\usepackage{epstopdf}
\usepackage{makeidx}
\usepackage{epsfig}
\usepackage{bm}
\usepackage{xcolor}
\usepackage[unicode=true, bookmarks=false,breaklinks=false,pdfborder={0 0 1},backref=false,colorlinks=false]{hyperref}
\usepackage{makeidx}
\hypersetup{colorlinks=true,citecolor=blue,linkcolor=blue,filecolor=blue,urlcolor=blue}

\newcommand{\be}{\begin{equation}}
\newcommand{\ee}{  \end{equation}}
\newcommand{\ba}{\begin{eqnarray}}
\newcommand{\ea}{  \end{eqnarray}}

\bibpunct{[}{]}{,}{n}{}{}

\begin{document}

\title{Electronic transport in disordered MoS$_2$ nanoribbons}

\author{Emilia Ridolfi}
\affiliation{Instituto de F\'{\i}sica,
Universidade Federal Fluminense, 24210-346 Niter\'oi, Brazil}

\author{Leandro R. F. Lima}
\affiliation{Instituto de F\'{\i}sica,
Universidade Federal Fluminense, 24210-346 Niter\'oi, Brazil}

\author{Eduardo R. Mucciolo}
\affiliation{Department of Physics, University of Central Florida, Orlando, FL 32816-2385, USA}

\author{Caio H. Lewenkopf}
\affiliation{Instituto de F\'{\i}sica,
 Universidade Federal Fluminense, 24210-346 Niter\'oi, Brazil}
 
\date{\today}

\begin{abstract}
We study the electronic structure and transport properties of zigzag
and armchair monolayer molybdenum disulfide nanoribbons using an
11-band tight-binding model that accurately reproduces the
material's bulk band structure near the band gap. We study the
electronic properties of pristine zigzag and armchair nanoribbons,
paying particular attention to the edges states that appear within the
MoS$_2$ bulk gap. By analyzing both their orbital composition and
their local density of states, we find that in zigzag-terminated
nanoribbons these states can be localized at a single edge for certain
energies independent of the nanoribbon width. We also study the effects of 
disorder in these systems using
the recursive Green's function technique. We show that for the zigzag
nanoribbons, the conductance due to the edge states is strongly
suppressed by short-range disorder such as vacancies. In contrast, the
local density of states still shows edge localization. We also show
that long-range disorder has a small effect on the transport
properties of nanoribbons within the bulk gap energy window.
\end{abstract}

\pacs{72.10.-d,72.80.Ga,73.23.-b,73.63.-b}

\maketitle

\section{Introduction}
\label{sec:introduction}

The wide interest in graphene has triggered an intense investigation
of the electronic and mechanical properties of other two-dimensional
materials \cite{wang2012, Geim2013, Butler2013}. Among them, transition metal dichalcogenides
(TMDs), and particularly molybdenum disulfide (MoS$_2$), are of great
appeal due to their finite band gap which could be explored for
optoelectronic applications \cite{wang2012}. Understanding the reasons
behind the poor mobility of the present-day state-of-the-art TMD
samples is a subject of intense theoretical and experimental debate 
\cite{Qiu2013, Yu2014, Kis2013, Schmidt2014, Baugher2013, Zhu2014, Ghatak2014}
 and a challenge for future practical uses of
these materials in devices.

Impressive advances in sample production have been
reported. Molybdenum disulfide nanowires and nanoribbons with
subnanometer width have been recently synthesized, with good quality
edges, mostly zigzag
terminated \cite{Li2005,LiuXiaofei2013_experimental,Wang2010,Xu2016,Koos2016}.
While great progress has been made on the experimental side, the
theoretical understanding of the properties of these systems is still
very limited. The theoretical literature consists mainly of density
functional theory (DFT) studies \cite{ li-yafei2008-dft, Botello-Mendez2009, Wang2010,
 Ataca2011,Gibertini2015, Pan2012, Yu2016, dolui2012,  kim2015-dft, 
 zhang2015-dft, Bollinger2001, Bollinger2003, chu2014-dft} 
that address only the electronic structure of pristine and narrow TMD
nanoribbons.

As in graphene, the presence of edges dramatically modifies the
low-energy spectrum of TMDs. DFT studies of MoS$_2$ find very distinct
features in the band structure of nanoribbons as compared with the
bulk: MoS$_2$ nanoribbons can be metallic depending on the orientation
of the edges. Zigzag nanoribbons typically show ferromagnetic and
metallic behavior, irrespective of their width, thickness, and passivation 
\cite{li-yafei2008-dft, Botello-Mendez2009,Wang2010,Ataca2011,Gibertini2015}. 
By increasing the nanoribbon width, it was found that the metallic 
edge state bands are preserved, 
{remaining as gap states (inside the bulk gap)}
but ferromagnetism is rapidly suppressed 
\cite{li-yafei2008-dft}. 
In contrast, Refs.~\cite{Pan2012} and \cite{Yu2016}
predict a semiconductor ($n$- or $p$-type) or half-metallic behavior
depending on the nanoribbon zigzag edge saturation. Most studies
\cite{Wang2010,Ataca2011,zhang2015-dft,dolui2012,kim2015-dft,
  li-yafei2008-dft, Pan2012} find that armchair nanoribbons are
nonmagnetic and semiconducting. To the best of our knowledge, the only
exception is Ref.~\cite{Botello-Mendez2009}, which reports
metallic armchair nanoribbons with a magnetic moment depending on the
passivation condition.

Although insightful, these results cannot be directly used to model
realistic TMD nanoribbons. The computational cost of DFT does not
allow one to address nanoribbons with realistic sizes or disorder, a
ubiquitous feature in systems synthesized these days. To account for
these limitations, one needs a computationally more efficient model
that accurately describes the MoS$_2$ low-energy bands and that is 
suitable for a disorder modeling at the atomistic scale. 
{
To study disorder effects in TMD nanoribbons, in this paper we consider a 
tight-binding model. Due to its single-particle nature, the tight-binding Hamiltonian 
does not describe the magnetic features of the nanoribbon edge states. 
On the other hand, we recall that the edge magnetization similar to TMDs has 
been extensively theoretically studied in pristine graphene nanoribbons
\cite{Yazyev2010,Carvalho2014}. 
The effect remains elusive to experiments and has been observed only
indirectly \cite{Tao2011}, possibly because edge disorder quenches magnetic 
properties \cite{Wimmer2008}. 
}

\begin{figure}[thbp]
        \centering
				\includegraphics[scale=0.7]{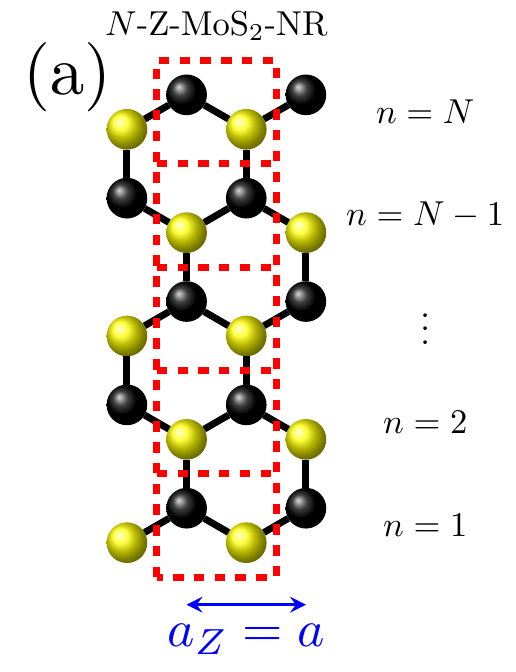}
				\includegraphics[scale=0.7]{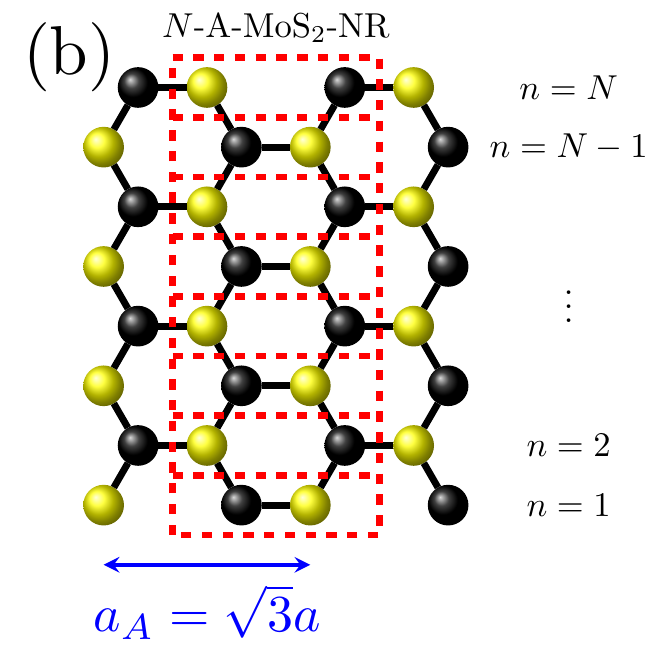}
 \caption{(Color online) Transverse unit cells of zigzag (a) and
   armchair (b) MoS$_2$ ribbons. The arrows indicate the lattice
   parameters $a_Z$ and $a_A$. The red dashed boxes mark zigzag lines
   (a) or armchair dimers (b) with its respective value $n$ indicated
   to the left. Dark (light) circles represent Mo (S) atoms. Note that
   in the zigzag nanoribbon one edge is S terminated while the
   opposite one is Mo terminated.}
 \label{fig:ribbons}
\end{figure}

One of the most used tight-binding implementations for MoS$_2$ \cite{chu2014-dft}
uses only the $d$ orbitals of
Mo. Hence, sulfur vacancies, one of the most important sources of
disorder, cannot be described by this model. A recent paper
\cite{Rostami15} uses an 11-band model to study the electronic
band structure of ribbons with about 30 nm of width. Reference
\cite{Rostami15} finds that nanoribbons with zigzag edges are
metallic, with edge states that close the bulk gap energy region,
while nanoribbons with armchair edges are semiconductors. This is in
line with recent scanning tunneling microscopy (STM) spectroscopy results \cite{Zhang2014,Koos2016}
that report the occurrence of a metallic phase within the bulk gap at
the (zigzag) edges of a MoS$_2$ monolayer on graphite. Similar results
have been also reported for MoS$_2$ on epitaxial
graphene \cite{Liu2016}. These experimental observations suggest that
the metallic edge modes are robust to disorder.

In this paper we use the tight-binding model put forward in
Ref.~\cite{Ridolfi2015} to systematically study the electronic
properties of monolayer MoS$_2$ nanoribbons. Our calculations
reproduce qualitatively the band structure of narrow nanoribbons
obtained by DFT. We show the necessity of considering the full
Hamiltonian, with even and odd parities with respect to $z$-axis
reflection, for an accurate description of nanoribbon electronic
states within the bulk gap energy window. By doing so, we observe the
appearance of an odd-parity band close to the Fermi level for both
kinds of edge terminations. In the zigzag case, we analyze the edge
nature of states inside the bulk gap. Interestingly, we find that the
metallic bands correspond to states localized at a single edge independent of 
nanoribbon width and disorder.

We also study the conductance and local density of states (LDOS) of
both pristine and disordered MoS$_2$ zigzag and pristine armchair
nanoribbons using the recursive Green's function method. We focus our
attention on the effect of short and long-range disorder on the
conductance of zigzag nanoribbons. The results are interpreted in
terms of topological invariants and their robust protection
against disorder. We find that even a modest concentration of
vacancies close to the edges can cause a large transmission
suppression, particularly within the bulk gap energy window. 
 In contrast, long-range
scattering does not have a significant effect on the conductance.

The paper is organized as follows. In Sec.~\ref{sec:bands} we present
the theoretical model and the band-structure calculations for pristine
MoS$_2$ nanoribbons with zigzag and armchair edges. In
Sec.~\ref{sec:transport} we discuss the electronic transport in clean
and disordered nanoribbons with Fermi energy in the vicinity of the
bulk gap. First, in Sec.~\ref{sec:pristine} we analyze the
conductance of pristine zigzag and armchair nanoribbons, supplemented
by a discussion of the LDOS of the zigzag case. Second, in Sec.
\ref{sec:unclean} we study the effect of both short- and long-range
disorder in the transmission and the LDOS of zigzag
nanoribbons. Finally, we draw some conclusions in
Sec.~\ref{sec:conclusions}.

\section{Band structure of pristine nanoribbons}
\label{sec:bands}

In this section we present the band-structure calculations for MoS$_2$
nanoribbons for both armchair and zigzag edges using the tight-binding
model introduced in Ref.~\cite{Ridolfi2015}.

Figure \ref{fig:ribbons} shows the different kinds of MoS$_2$
nanoribbon unit cells considered in this paper and serves a guide for
the notation. We consider zigzag (Z-MoS$_2$-NR) and armchair MoS$_2$
nanoribbons (A-MoS$_2$-NR) with translational invariance along the
``horizontal'' direction and a finite width along its ``vertical''
direction. Figure~\ref{fig:ribbons} shows the ribbons from a top view,
where two sulfur (S), one above and one below the plane containing the
molybdenum (Mo) atoms, sit on top of each other. For notation
convenience, we identify ribbons with different widths as
$N$-Z-MoS$_2$-NR and $N$-A-MoS$_2$-NR, where the integer $N$
corresponds to the number of zigzag lines and the number of armchair
dimers, respectively, indicated by the dashed
rectangles \cite{Wakabayashi2012}.
Each Mo (S) atom contains five (three) orbitals corresponding, in the
limit $N\gg 1$, to a bulk MoS$_2$ unit cell with three atoms (one Mo and
two S) with a total of $11$ orbitals \cite{Ridolfi2015}. Since both
zigzag lines and armchair dimers have one Mo and two S atoms, the
total number of orbitals in any ribbon is $11N$. We define the zigzag
and armchair lattice parameters as $a_{Z}=a$ and $a_{A}=\sqrt{3}a$,
respectively, where $a=3.16$ \AA\ is the Mo-Mo distance.

As discussed in Ref.~\cite{Ridolfi2015}, the full model
Hamiltonian for the MoS$_2$ monolayer can be decoupled into odd and
even symmetry parts with respect to the Mo plane ($z$ symmetry). For a
MoS$_2$ monolayer, the orbital compositions of the valence band and
the conduction band are mostly even \cite{Ridolfi2015,Rostami15}. In
Ref.~\cite{Rostami15}, the authors take advantage of this fact
and compute the band structure of MoS$_2$ nanoribbons using only the
even part of the Hamiltonian, asserting that odd parity bands are
energetically far away from the bulk gap. Unfortunately that is not
the case for MoS$_2$ nanoribbons, as we show in the following.

\begin{figure}[htbp]
	\centering
				\includegraphics[width=0.45\columnwidth]{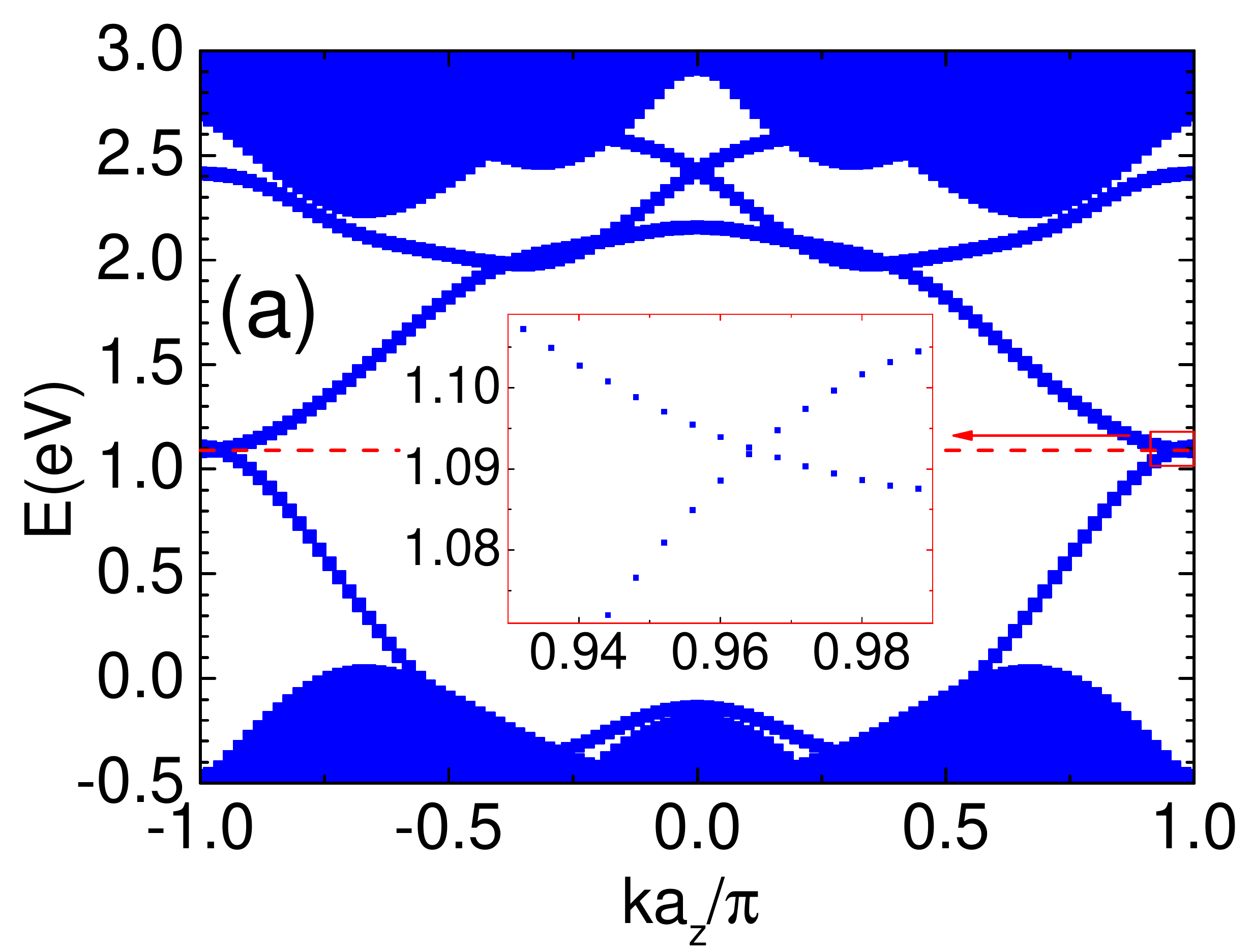} 
				\includegraphics[width=0.45\columnwidth]{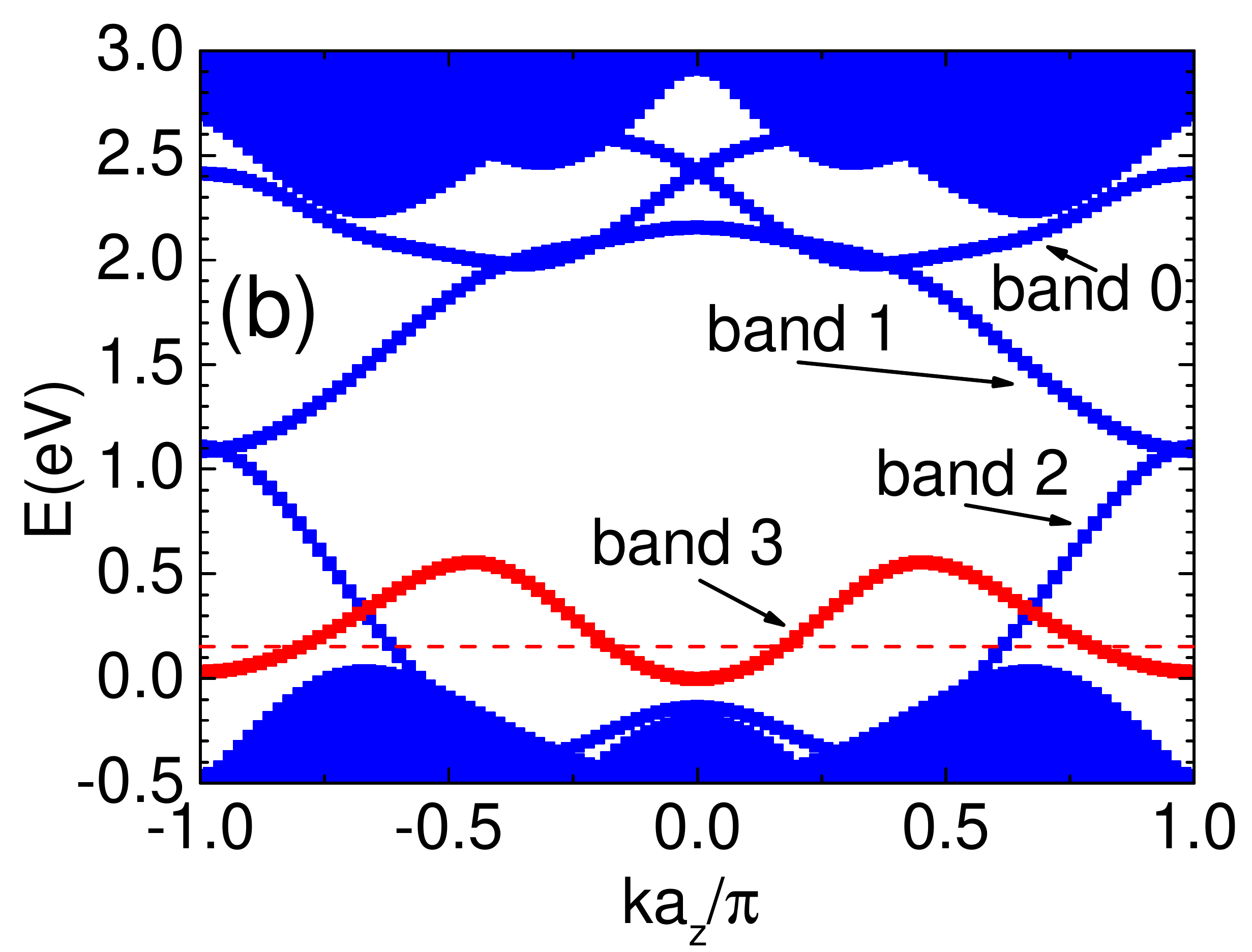}
				\includegraphics[width=0.45\columnwidth]{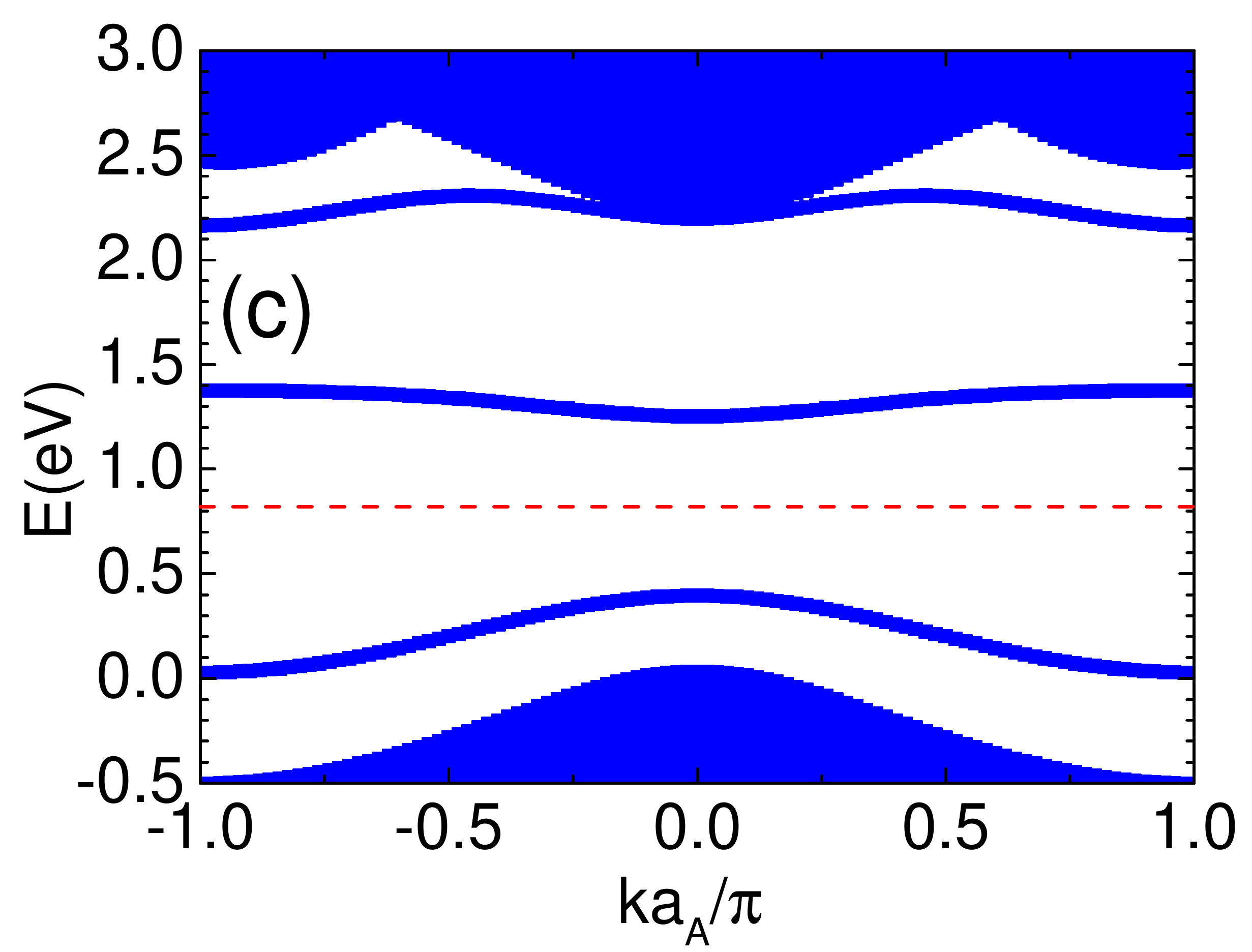} 
				\includegraphics[width=0.45\columnwidth]{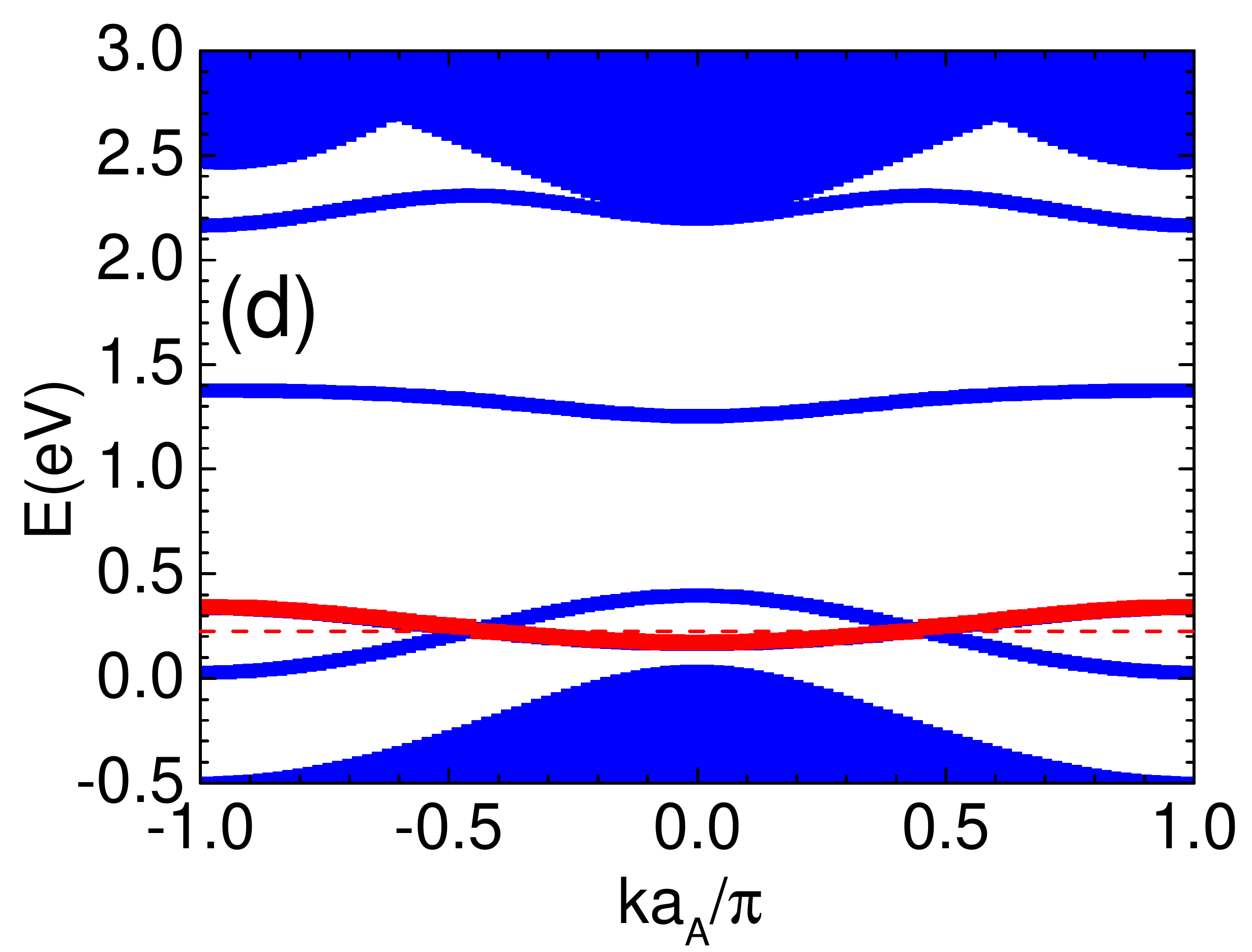}
 \caption{(Color online) {Spin-unpolarized} electronic band structure of
   nanoribbons with different edge orientations: (a) Band structure of
    90-Z-MoS$_2$-NR for the even-band model with $540$ orbitals, with
   the inset showing the level crossing that takes place near the
   Brillouin zone edge. (b) Band structure for the all-band model with
   $990$ orbitals, where we identify zigzag midgap states as the four
   bands (from $0$ through $3$). The bottom figures show the band
   structure of the 90-A-MoS2-NR using the even-band model with $540$
   orbitals (c) and the all-band model with $990$ orbitals (d). The
   odd parity bands are plotted in red. The band structures in (c) and
   (d) are doubly degenerate. {The red dashed line represents the Fermi level.}}
 \label{fig:bands}
\end{figure}

Due to finite-size effects, the band structure of MoS$_2$ nanoribbons
can be significantly different from that of bulk monolayers. The bands
suffer reorganization and/or hybridization which can enhance the
contribution of the odd symmetry bands in the bulk gap region. For this
reason we implement the all-band model (ABM), consisting of even and
odd symmetry bands, to compute the band structure of Z-MoS$_2$-NR and
A-MoS$_2$-NR. Those are compared with the even-band model (EBM). We
study ribbons characterized by widths comparable with experimentally
produced samples \cite{Li2005, LiuXiaofei2013_experimental,
  Wang2010, Xu2016,Koos2016}.

Typical results are shown in Fig.~\ref{fig:bands}. As predicted
earlier \cite{Rostami15}, clean zigzag ribbons are metallic. 
This applies for both the even- and the all-band models.
{
In contrast, pristine armchair ribbons are semiconductor for the
even-band model \cite{Rostami15}  and metallic when even and odd bands 
are considered (this work). 
In neither case do the edge state bands extend over the whole
energy window corresponding to the bulk gap above the Fermi level.
Consequently, armchair MoS$_2$ nanoribbons with a small electron doping
\footnote{Taking spin degeneracy into account,
this corresponds to two electrons in excess over the whole nanoribbon
area.}
also become semiconductor in the all-bands model.
}

The comparison between 
the two models unravels the presence of an odd band around the energy of 
$0.25$ eV
for both edge orientations, as highlighted in red in
Figs.~\ref{fig:bands}(b) and \ref{fig:bands}(d). We note that in the
armchair case the bands in the energy interval $0 \lesssim E
\lesssim 2.25$ eV of Figs.~\ref{fig:bands}(c) and \ref{fig:bands}(d)
are roughly doubly degenerate, while in the zigzag case
[Figs.~\ref{fig:bands}(a) and \ref{fig:bands}(b)] there is no
degeneracy.

\begin{figure}[htbp]
	\centering
				\includegraphics[width=0.45\columnwidth]{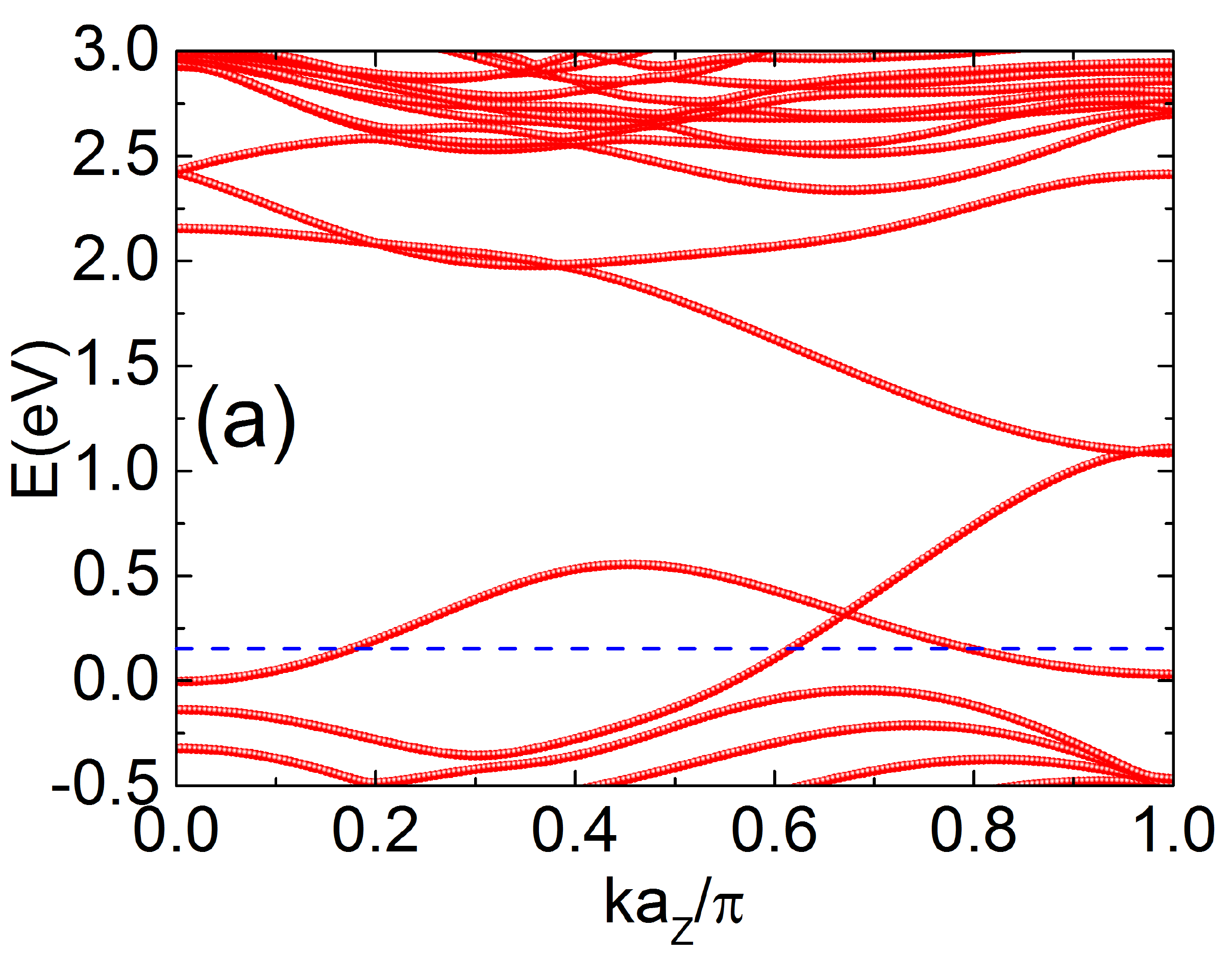} 
				\includegraphics[width=0.45\columnwidth]{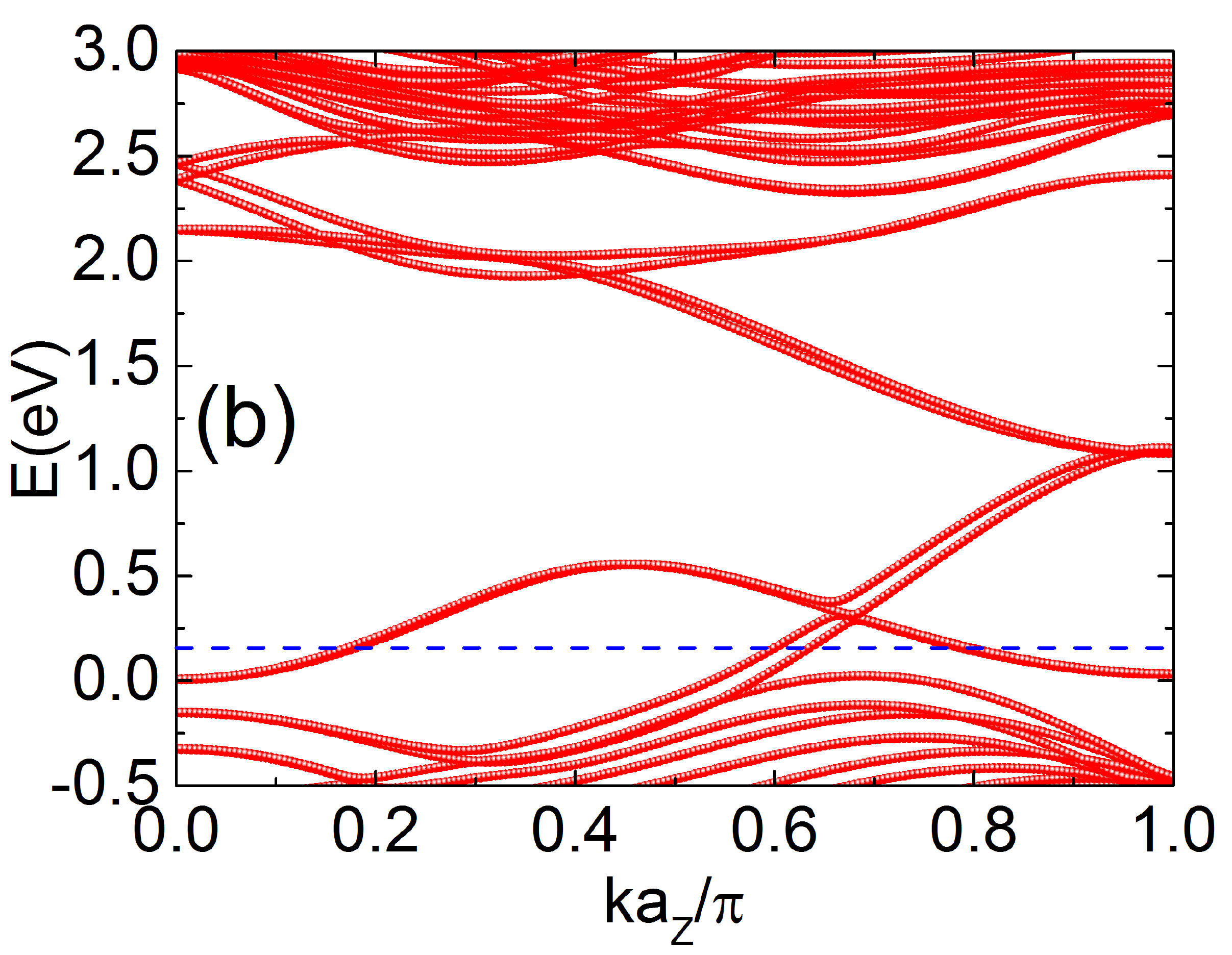}
				\includegraphics[width=0.45\columnwidth]{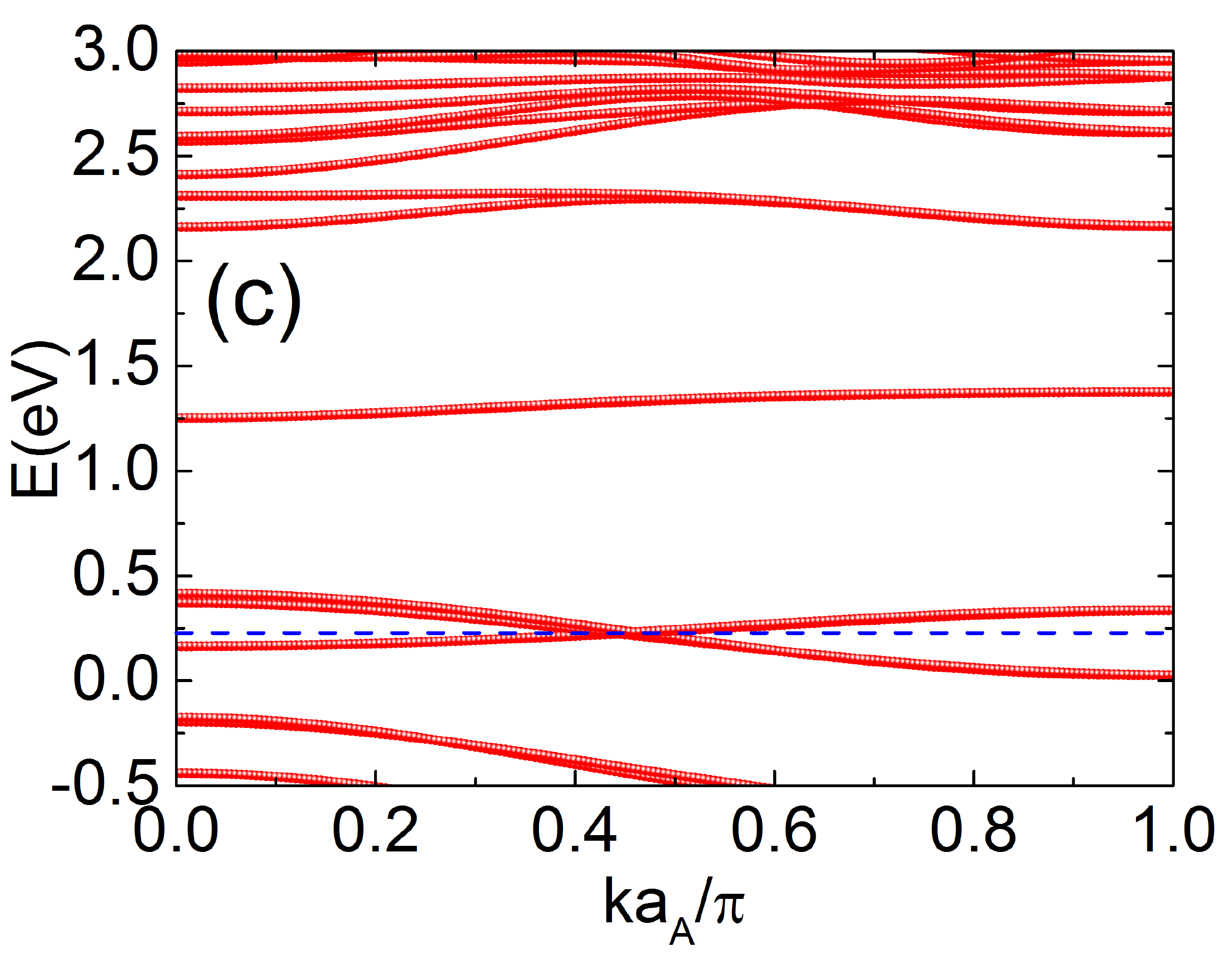} 
				\includegraphics[width=0.45\columnwidth]{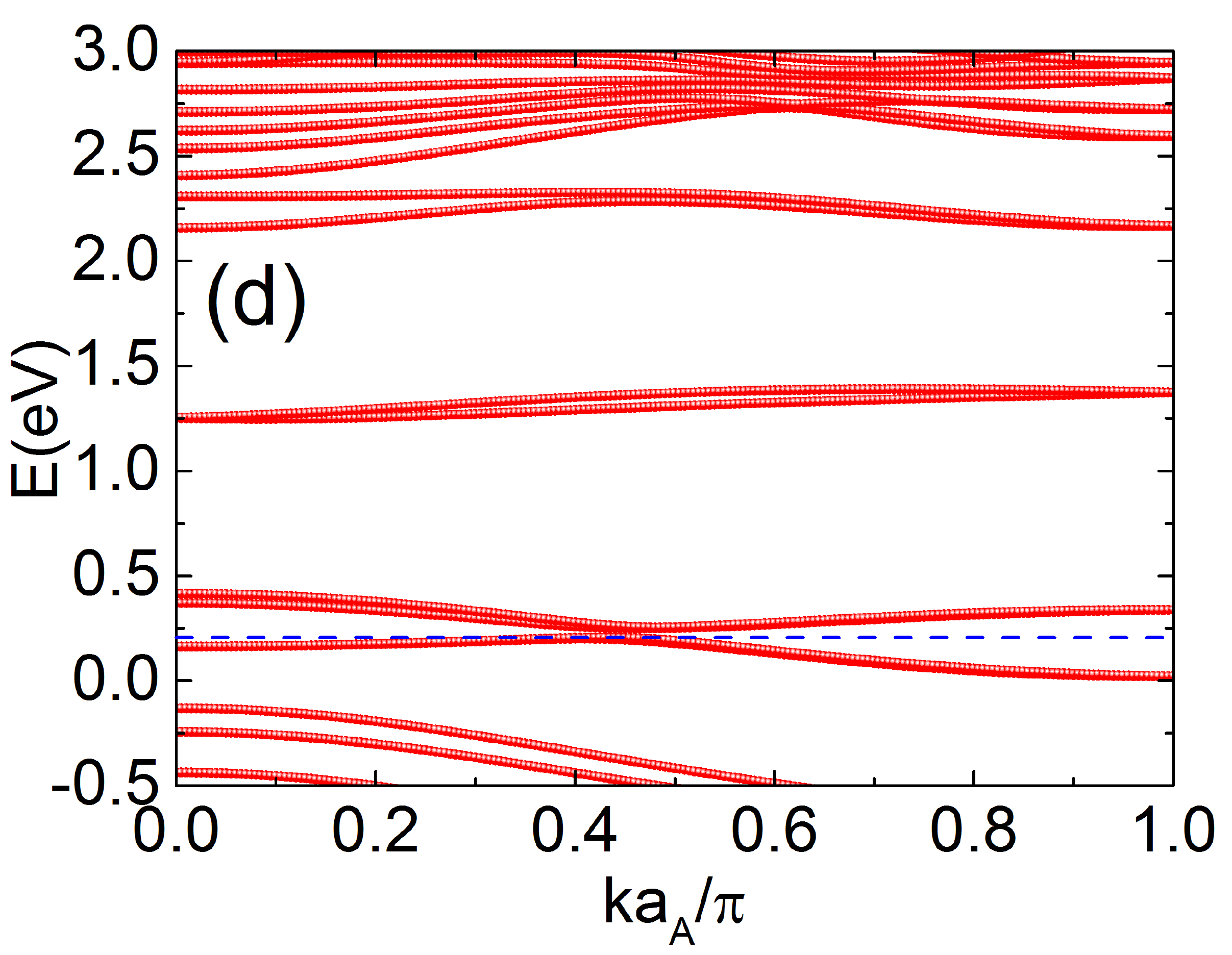}
 \caption{(Color online) Comparison between the {spin-unpolarized} and the
   {spin-polarized} band structures of zigzag and armchair MoS$_2$ nanoribbons.
   The top panels correspond to the band structure of the
   10-Z-MoS$_2$-NR with (b) and without (a) spin resolution within the
   all band-model with 110 orbitals. Analogously, the bottom panels
   show the band structure of the 10-A-MoS$_2$-NR with (d) and without
   (c) spin resolution. {The blue dashed line represents the Fermi level.}}
 \label{fig:spinbands}
\end{figure}

A quantitative comparison between tight-binding and DFT band structure
calculations for nanoribbons is difficult, since most DFT studies
assume edge passivation conditions which are not included in the
tight-binding model. Moreover, DFT accounts for electron-electron
interactions due to the non-homogeneous charge distribution at the
vicinity of the nanoribbon edges. Nonetheless, the band structures we
obtain show good qualitative and quantitative agreement with results
from the DFT literature \cite{Wang2010, li-yafei2008-dft,
  zhang2015-dft, Bollinger2001, Bollinger2003,
  Botello-Mendez2009,dolui2012,Ataca2011,chu2014-dft,
  kim2015-dft,Pan2012, Yu2016}. The comparison with the band structure
presented in Ref.~\cite{Andersen2014} for zigzag nanoribbons
where the authors find a band very similar to our ``odd band'' is
particularly reassuring.

The band structures presented so far are not spin resolved. We include
spin resolution following the prescription reported in
Ref.~\cite{Ridolfi2015}. We consider the all-band model with
$11$ orbitals per bulk unit cell and introduce the spin-orbit coupling
term $H_{SO}=\sum_\alpha \frac{\lambda_\alpha}{\hbar^2}\mathbf
L_\alpha \cdot \mathbf S_\alpha$. Here, $\lambda_\alpha$ is the
intrinsic atomic spin-orbit strength for $\alpha=$Mo or S, $\mathbf
L_\alpha$ is the atomic orbital angular momentum operator, and
$\mathbf S_\alpha$ is the electronic spin operator acting on all atoms
of the system. We set the parameters $\lambda_{\rm Mo}=75$ meV and
$\lambda_{\rm S}=52$ meV \cite{Ridolfi2015,Roldan14}, which reproduce
the experimental spin splitting at the $K$ point of the bulk MoS$_2$
valence band \cite{Miwa2015}.

In Fig.~\ref{fig:spinbands} we show the band structure for zigzag and
armchair MoS$_2$ nanoribbons with and without the spin-orbit
interaction in the energy window of interest. The spin-resolved bands
of the 10-Z-MoS$_2$-NR and the 10-A-MoS$_2$-NR in
Figs.~\ref{fig:spinbands}(b) and Fig.~\ref{fig:spinbands}(d),
respectively, show that the main bands in the interval $0
\alt E \alt 2.5$ eV present very small spin splittings. (For some bands
the effect is hard to notice within the resolution of the figures.)
We use the full spin-orbit coupling Hamiltonian, including both spin-conserving and spin-flipping terms. We have verified that the spin-splitting effect on the band structure is due to the diagonal
(spin-conserving) terms of the Hamiltonian, while off-diagonal
(spin-flipping) terms give a negligible contribution. In the bulk
region, for energies $E<0$ eV and $E>2.5$ eV, the spin splitting is
more pronounced.
We find that including spin-orbit coupling in our all-band model
provides a spin splitting in the armchair band structure in
Fig.~\ref{fig:spinbands}(d) which is not found in
Ref.~\cite{Rostami15}.

The small energy splitting of the spin-resolved bands suggests that we
can safely neglect the intra-atomic spin-orbit coupling and treat the
system as spin degenerate when addressing transport properties.

Let us now focus our discussion on the orbital character of the
metallic bands of zigzag nanoribbons in Fig.~\ref{fig:bands}(b).
Reference~\cite{Rostami15} established the edge nature of two of
the even bands inside the gap (corresponding to bands 1 and 2 in
Fig.~\ref{fig:bands}) and the delocalized nature of the valence and
conduction bands by plotting their wave functions across the ribbon
width. In our model, bands $1$ and $2$ touch each other very close to
the Brillouin zone (BZ) edge at $k_c=\pm 0.96\pi/a_Z$, close to the
bulk $K$ and $K'$ points, having 
{an energy difference of roughly $20$ meV at the BZ edge. Those features 
are nearly imperceptible in Fig.~\ref{fig:bands}(b). 
In Ref.~\cite{Rostami15} the bands cross further away from the BZ edge, at  
$k_c \approx \pm 0.87\pi/a_Z$, forming one-dimensional analogs of Dirac cones 
\cite{CastroNeto09}. The latter are more pronounced than the ones we obtain, 
with an energy difference of roughly $400$ meV at the BZ edge. We stress that these 
effects are purely orbital and each band is spin degenerate.}

We determine the edge nature of the bands from $0$ through $3$ in
Fig.~\ref{fig:bands}(b) by analyzing their wave functions along the
Brillouin zone.
By studying the squared wave functions of all atoms in the ribbon unit
cell (not shown here), we determine whether a particular state is
distributed mostly near the edges (edge state) or across the whole
ribbon (bulk state). We find that all midgap states in
Fig.~\ref{fig:bands}(b) are localized at the nanoribbon edges. Band
$1$ is located at the S-terminated edge while bands $0$, $2$ and $3$
are located at the Mo-terminated edge.
We also notice that one of the valence bands at around $-1$ eV has the
same edge nature as the bands $0$, $2$, and $3$ (it can be seen in
Fig.~\ref{fig:one_vacancy}, where a larger energy interval is shown),
while other usual bulk bands have Gaussian-like envelopes as noted in
Ref.~\cite{Rostami15}.

The appearance of metallic bands in MoS$_2$ ribbons can be interpreted
in topological terms. The first classification of the topological
properties of a system can be done according to its Chern
number \cite{Hasan10}. The latter distinguishes a simple insulator,
with Chern number equal to zero, from a system with topologically
protected edge states, equivalent to a quantum Hall state, where the
Chern number is non zero. This classification fails in systems with
time-reversal symmetry, since it leads to a zero Chern number even for
systems with topologically protected edge states \cite{Kane05,Hasan10}.
In these cases, it is customary to use another topological invariant,
the $Z_2$ invariant, that can be either $1$ (strong topological
insulator) or $0$ (trivial insulator or weak topological
insulator) \cite{Hasan10}.

The zigzag MoS$_2$ ribbon has $Z_2=0$, since the number of Kramers
pairs at a single edge is always even \cite{Kane05,Hasan10} [see
Fig.~\ref{fig:spinbands}(b)]. According to this classification, the
system behaves like an ordinary metal, namely, any kind of disorder has 
a strong effect on the system electronic transport properties.

However, if one considers a single valley and a single spin, the
system has a nontrivial topology. Both DFT \cite{Rostami15} and
tight-binding \cite{Rostami15} calculations obtain a large value for
the Berry curvature \cite{Hasan10} near the $K$ and $K^\prime$ points
of MoS$_2$ (bulk) monolayers. The study of the topology around these
points \cite{Rostami15} shows a non zero Chern number
$C_K=C_{K\downarrow}+C_{K\uparrow}$ with $C_{K'} =-C_K$. Thus, the
system has $C=C_K + C_{K'} = 0$. In other words, instead of the spin
texture of a two-dimensional topological insulator, MoS$_2$ nanoribbon
edge states exhibit a ``valley texture''. This nontrivial bulk-edge
correspondence results in the edge band state crossing shown in the
inset of Fig.~\ref{fig:bands}(a). The low-energy states of the zigzag
MoS$_2$ ribbon around the $K$ point, formed by the crossing of 
bands $1$ and $2$ (see Fig.~\ref{fig:bands}), behave as a quantum
valley Hall state \cite{Rostami15}. Hence, in the absence of disorder
sources that cause spin-flip transitions, the system is protected
against intra valley scattering processes.

Another way of understanding the ``local'' nature of the protection is
under the light of intervalley scattering
events \cite{Wakabayashi07,Wakabayashi09,Lima12}.
Away from the band crossing energy, states with opposite propagation
direction belong to different valleys. Thus, it is only possible to
enable backscattering (change the propagation direction) by
introducing a disorder source that provides a large enough amount of
momentum to scatter the electron state from one valley to the other.
Hence, depending on the value of $\Delta k$, backscattering can have a
large cross section in the presence of short-range scatterers, while
as a rule, it is negligible for long-range
ones \cite{Wakabayashi07,Wakabayashi09,Lima12}. We study these
processes extensively in the following sections.

{Before proceeding to the analysis of the transport properties, a comment on 
different tight-binding parametrizations for MoS$_2$ is in order. 
A recent study \cite{Gibertini2015} argues that the origin of the metallic states in 
zigzag nanoribbons is due to polar discontinuities at the edges. That paper
reports that DFT calculations give spin-polarized bands 
closing the bulk gap, while a Wannier tight-binding model leads to edge state 
bands with a band gap of about 0.3 eV, independent of the nanoribbon width. 
Based on these results, the authors \cite{Gibertini2015} conclude that metallicity
is driven by the electric fields caused by charge polarization at the system edges. 
In contrast, our tight-binding model (and also that of Ref.~\cite{Rostami15})
gives metallic edge state bands solely due to geometric effects, since we do not 
account for electronic interactions. 
We recall that the tight-binding model we use is not the same as the one of Ref.~\cite{Gibertini2015}.
In our case, the tight-binding parameters were obtained from an accurate fit of the band energies and
orbital compositions of MoS$_2$ bulk DFT calculations \cite{Ridolfi2015}.
In any event, despite the differences in spin polarization, both models indicate that
the edge states remain metallic, closing the bulk gap, irrespective of the nanoribbon width.
The understanding of the nature of the discrepancies is still unclear and beyond the scope of this paper,
which is the study of disorder effects on the conductance of MoS$_2$ nanoribbons
using a state-of-the-art single-particle effective Hamiltonian.
}

\begin{figure}[htbp]
\centering
\includegraphics[width=0.8\columnwidth]{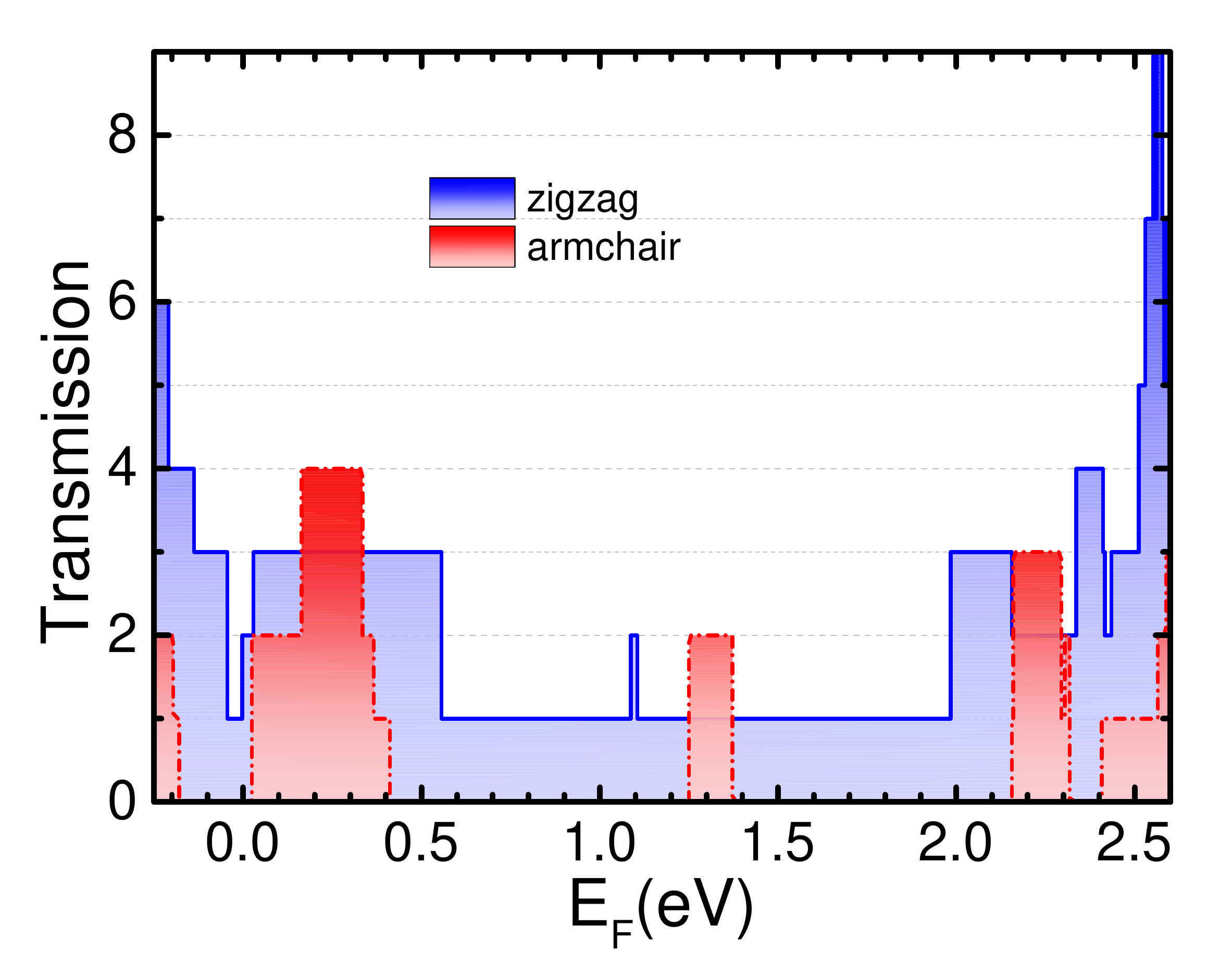}
\caption{(Color online) Transmission of pristine ribbons as a function
  of the Fermi energy. The solid blue curve is the transmission for
  10-Z-MoS$_2$-NR and the dashed red curve corresponds to 10-A-MoS$_2$-NR.}
 \label{fig:clean}
\end{figure}

\section{Transport in pristine and disordered nanoribbons}
\label{sec:transport}

In this section we study the electronic transport in MoS$_2$
nanoribbons. In the zigzag case, we discuss our findings in light of
the closing of the gap by the metallic edge states. We consider the
electronic transport in a system composed by a central part connected
to two electron reservoirs maintained at different chemical potentials
that act as source and drain terminals. The central part is a
nanoribbon of width $N$ (see Fig.~\ref{fig:ribbons}) and length $M$,
corresponding to the number of longitudinal unit cells (slices). We
calculate the Landauer conductance, given in terms of the electronic
transmission between the source (left reservoir $L$) and the drain
(right reservoir $R$), using the Caroli formula
\cite{Caroli1971,Datta1996},
\begin{eqnarray}
T_{RL}(E) =
\text{Tr}[\Gamma_{R}(E)G_{RL}^{r}(E)\Gamma_{L}(E)G_{LR}^{a}(E)]
\label{eq:caroli},
\end{eqnarray}
where $\Gamma_{L}$($\Gamma_{R}$) is the decay width function of left
(right) contact, and $G^{r}_{RL}$ ($G^{a}_{LR}$) is the retarded
(advanced) Green's function describing the propagation from $L$ to $R$
(from $R$ to $L$).

We model the contacts as semi-infinite lattices, calculate the level
width functions $\Gamma_L$ and $\Gamma_R$ using the decimation
technique \cite{Sancho85,Lewenkopf2013}, and $G^{r}_{RL}$ using the
recursive Green's function method (RGF) \cite{MacKinnon85}. This method
allows for a numerically efficient computation of the electronic
transport properties and spectral properties \cite{Lewenkopf2013} such
as the local density of states, namely
\begin{align}
\rho(\mathbf r_{\alpha j},E) = -\frac{1}{\pi}\text{Im}[G_{\alpha j,\alpha j}^{r}(E)].
\label{LDOS}
\end{align}
Here, $j$ labels the orbital inside the atom $\alpha$. It also allows
for an amenable inclusion of several disorder mechanisms at the
microscopic level \cite{Lewenkopf2013}.

In the following we present our numerical calculations of the
electronic transmission and LDOS for pristine nanoribbons and for
zigzag nanoribbons with defects.

\subsection{Pristine zigzag and armchair nanoribbons}
\label{sec:pristine}

In Fig.~\ref{fig:clean}, we present the transmission $T_{RL}(E)$ for
pristine MoS$_2$ nanoribbons. The results correspond to nanoribbons
of widths comparable to the ones considered in the band-structure DFT
literature \cite{Wang2010, li-yafei2008-dft,
  zhang2015-dft, Bollinger2001, Bollinger2003, Botello-Mendez2009,
  dolui2012,Ataca2011, chu2014-dft, kim2015-dft, Pan2012}. We note that
these nanoribbons have smaller widths than the ones considered in the
band-structure study of the previous section. Nonetheless, we have
checked that their metallic edge states remain almost invariant with
increasing width. Thus, the chosen ribbon width allows for faster
computation of the electronic transport quantities and still gives
insight on the behavior of the edge modes of the wider ribbons.

Figure \ref{fig:clean} shows the pristine ribbon transmission as a
function of the Fermi energy $E_F$. The transmission is quantized and
consistent with the number of available energy bands as a function of
$E_F$. As expected, the zigzag edges give origin to metallic edge
states (or transmission channels) that make the conductance nonzero
over the whole energy interval corresponding to the bulk gap. In
contrast, the energy band for nanoribbons with armchair edges splits
the transport gap into two smaller ones.

\begin{figure}[htbp]
        \centering
        \includegraphics[width=0.80\columnwidth]{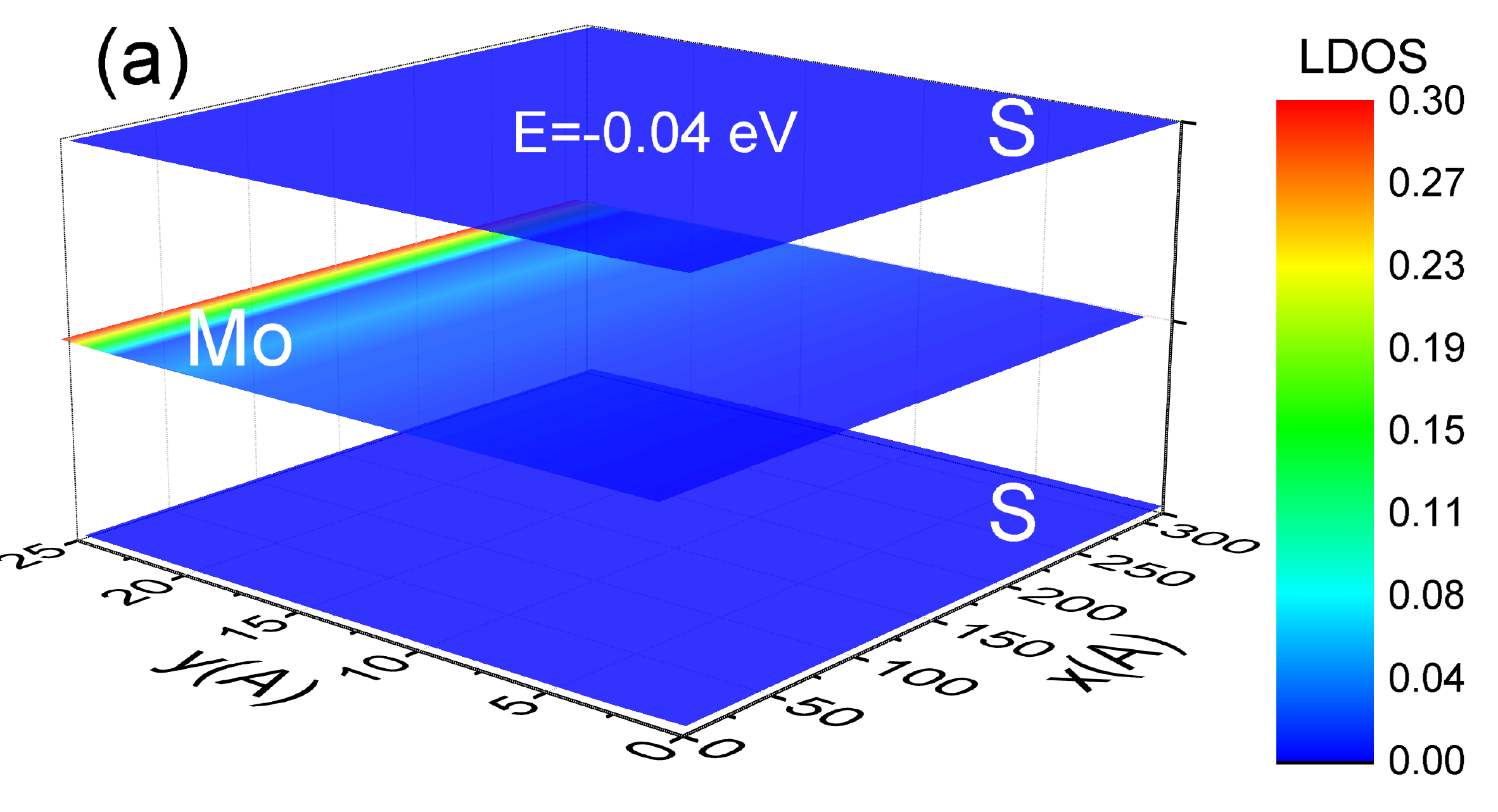}
				\includegraphics[width=0.80\columnwidth]{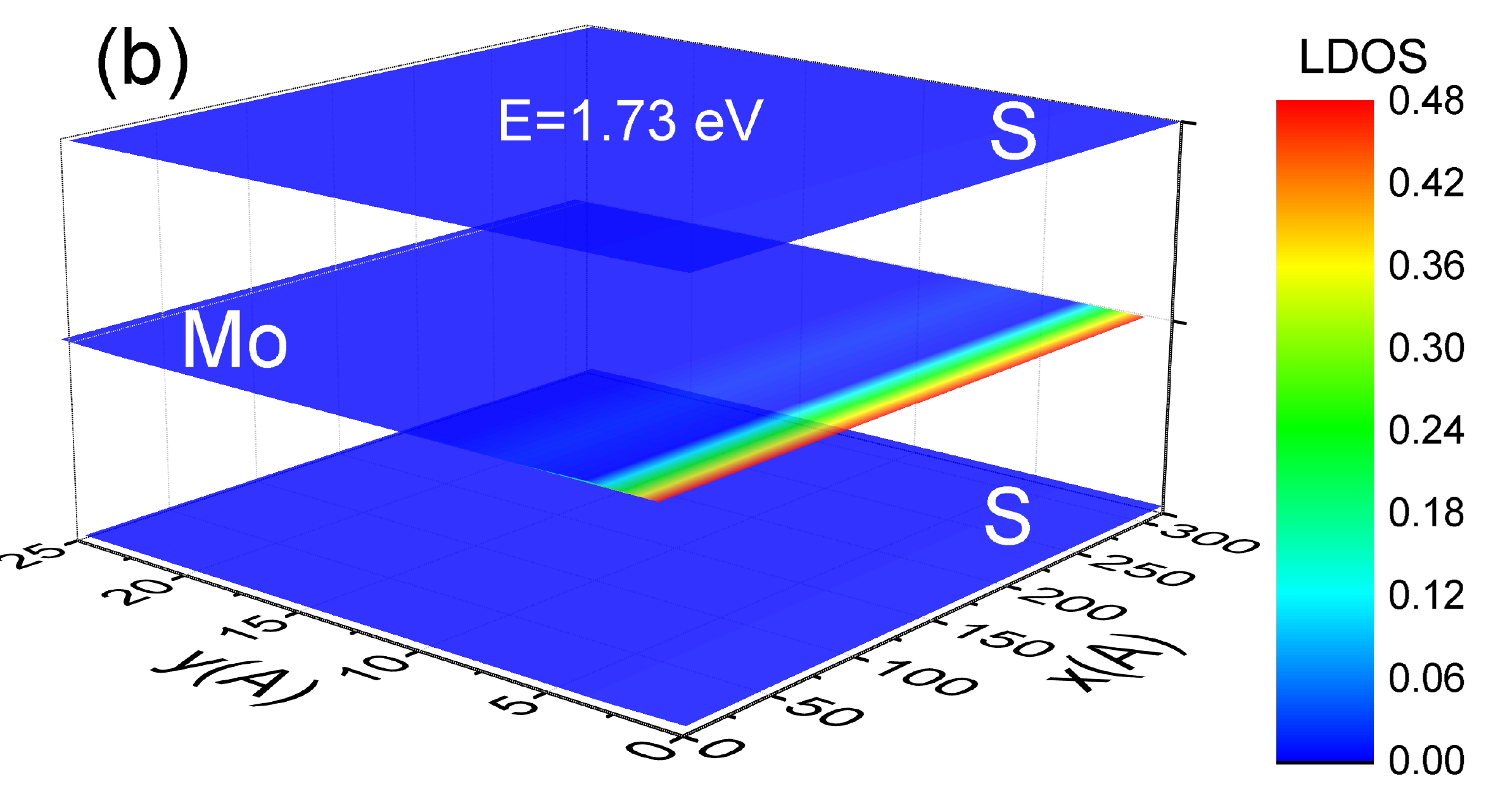}
				\includegraphics[width=0.80\columnwidth]{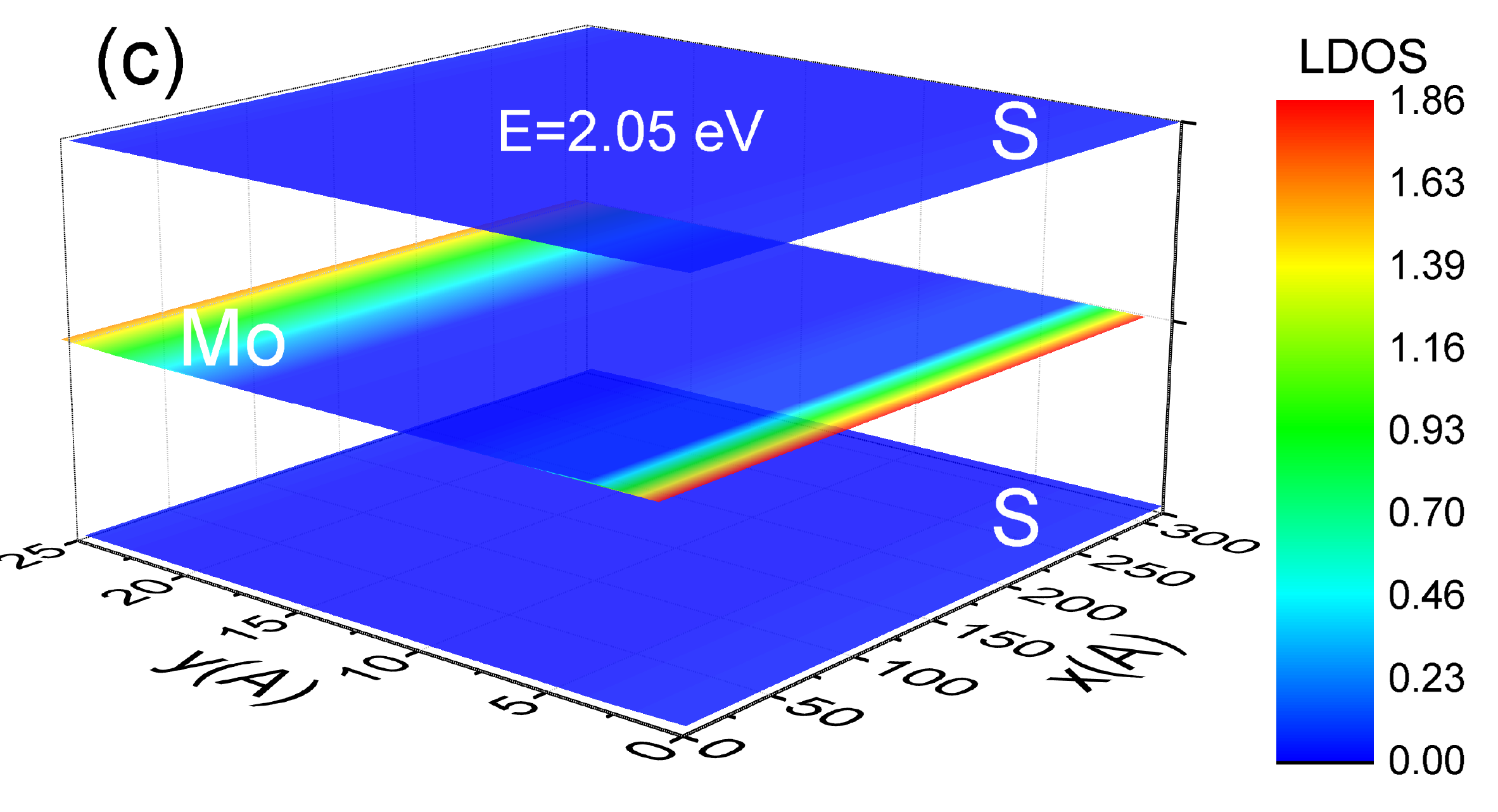}
				\includegraphics[width=0.80\columnwidth]{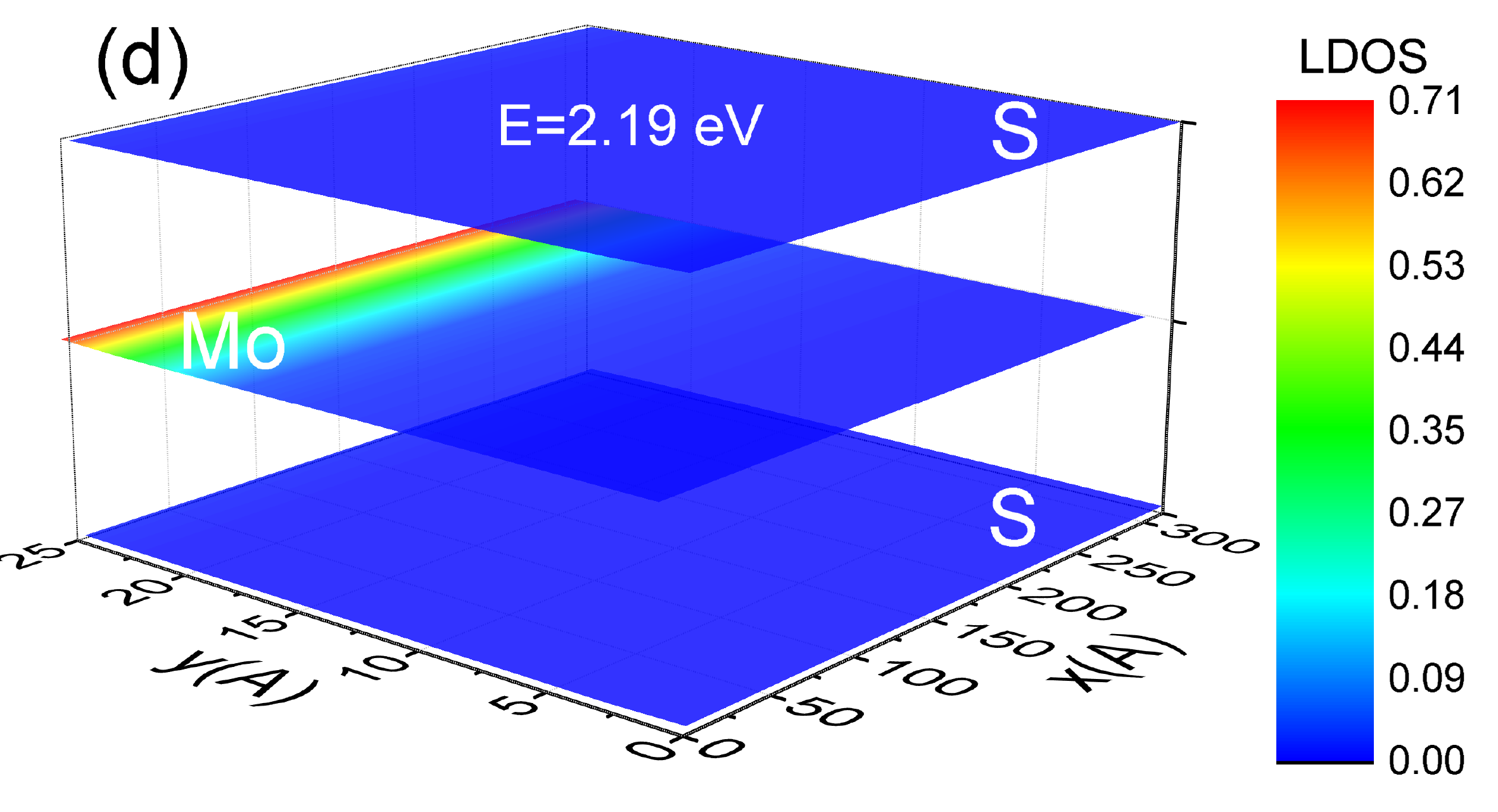}
				\includegraphics[width=0.80\columnwidth]{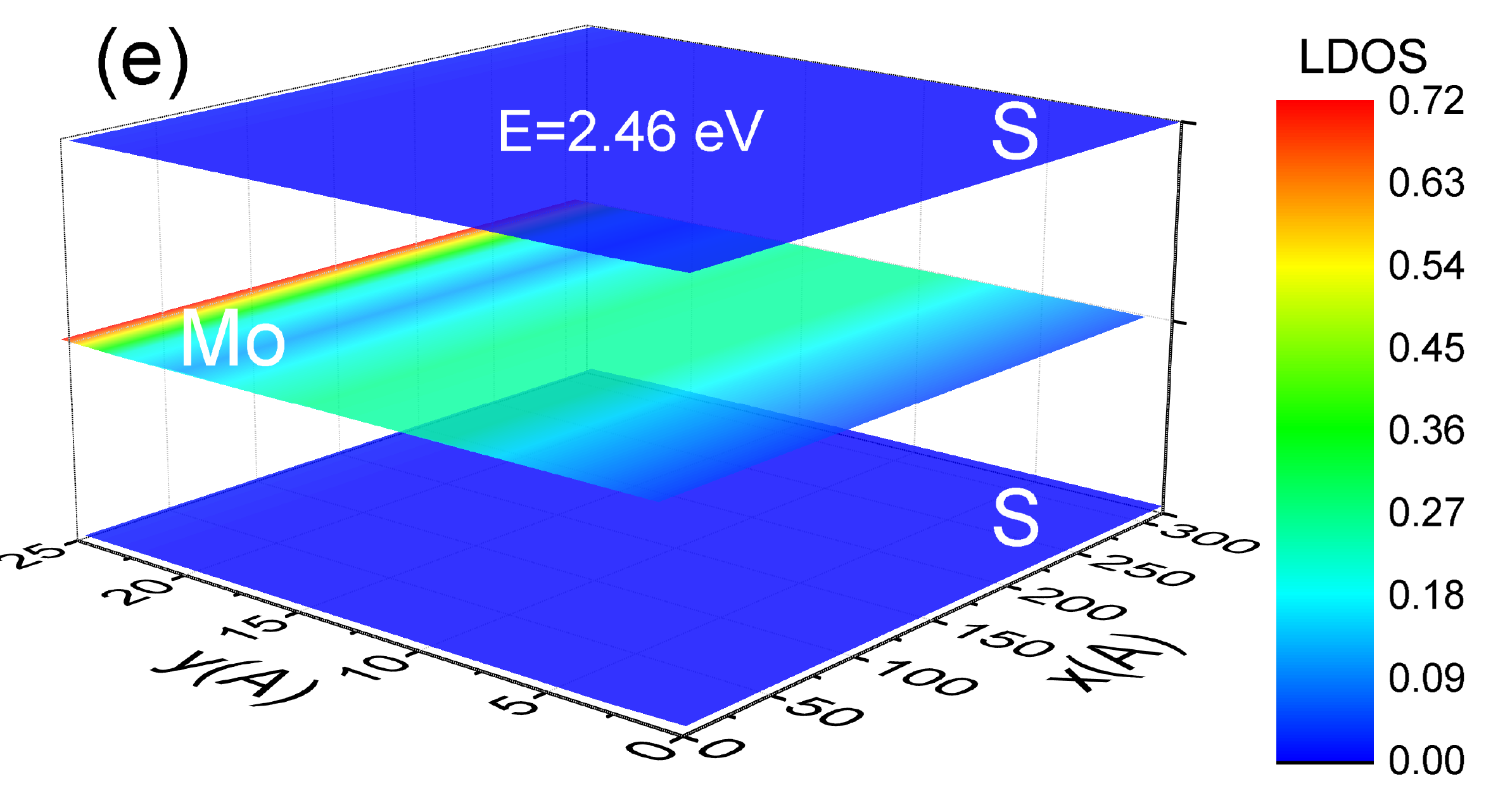}
 \caption{(Color online) LDOS at energies (a) $E=0.04$ eV, (b)
   $E=1.73$ eV, (c) $E=2.05$ eV, (d) $E=2.19$ eV, and (e) $E=2.46$ eV
   for the pristine 10-Z-MoS$_2$-NR. The ribbon edge at $y=0$ is
   S-terminated while the opposite one is Mo-terminated. The image in
   each plane corresponds to a single atomic nature, S or Mo, as
   indicated by the labels. In (a) bands $2$ and $3$ contribute with
   available states, while in (b), (c), and (d) only states from bands
   $2$, $1$, and $0$ contribute, respectively. In (e) the conduction
   bulk states are present across the whole ribbon width.}
 \label{fig:LDOS_clean}
\end{figure}

Figure \ref{fig:LDOS_clean} shows the LDOS of the zigzag MoS$_2$
nanoribbon. In the considered geometry, the edge located at $y=0$ is
terminated by S atoms while the opposite edge at $y \approx 25.5$
\AA\ is terminated by Mo atoms, as indicated by the labels in
Fig.~\ref{fig:LDOS_clean}. We find that the orbital composition of all
metallic bands is dominated by $d$ orbitals located at the Mo layer
with a very small LDOS at the S layers. The states are localized at a
single nanoribbon's edge, which we denote according to its termination
(either S or Mo). We choose five representative energies on the main
plateaus of Fig.~\ref{fig:clean} corresponding to metallic edge
states. The edge states belonging to band $1$ in
Fig.~\ref{fig:LDOS_clean}(c) $E=1.73$ eV are located at the
S-terminated edge. The other bands $0$, $2$, and $3$ have states
mainly distributed at the Mo-terminated edge [see
Fig.~\ref{fig:LDOS_clean}(b) $E=0.82$ eV and \ref{fig:LDOS_clean}(d)
$E=2.19$ eV]. Interestingly, Fig.~\ref{fig:LDOS_clean}(c) shows a LDOS
distributed at both nanoribbon edges, since both bands $0$ and $1$
contribute to the LDOS at energy $E=2.05$ eV. Figures
\ref{fig:LDOS_clean}(a) and \ref{fig:LDOS_clean}(e) show the LDOS for
$E=-0.04$ eV and $E=2.46$ eV, respectively. These energies are outside
the ``bulk energy gap'' window, but the corresponding states are still
influenced by metallic bands. 
Due to this fact, the states are mostly localized at the Mo-terminated
edge, but a bulk character is manifest.

\subsection{Zigzag nanoribbon with defects}
\label{sec:unclean}

Motivated by the topological nature of the metallic bands, we study the
robustness of these states around the valleys $K$ and $K'$ against
short- and long-range disorder.

Several kinds of defects in MoS$_2$ have been addressed in the
literature, such as vacancies, adatoms, substitutional doping,
structural and topological defects, folding, wrinkling, and
rippling \cite{Lin2016}. Among the zero-dimensional defects, the most
common are single S vacancies \cite{Lin2016}. In this study we also
consider double S vacancies and Mo vacancies without
reconstruction. As for long-range defects, the most studied ones in
MoS$_2$ are ripples \cite{Meyer2007,Brivio2011,Luo2015}. These deform
the bonds and modify the interatomic distance of the nanoribbon
structure. As a consequence, in a realistic model, the on-site and the
hopping parameters of the tight-binding model need to be modified. In
addition, electronic charges trapped in the substrate also unavoidably
provide a source of local potential disorder. In this paper we address
the simple case of local potential scattering for the long-range
potential disorder case.

\subsubsection{Short-range scattering}

Let us study how robust are the edge states in Z-MoS$_2$-NR against
short-range disorder, such as vacancies which are ubiquitous in
MoS$_2$ \cite{Lin2016}. Single and double sulfur (top and bottom)
vacancies are the ones with the lowest energy formation \cite{Lin2016}. We
also consider single Mo vacancies as a model case.

We start with the simplest case of a single short-range defect. We
model the defect as a single vacancy by adding a large on-site
potential at the lattice atom of interest. We have numerically
verified that the large on-site vacancy model is fully equivalent to
cutting all the hopping connections between the orbitals of the
vacancy atom and the other atoms in the lattice
\footnote{We find that this condition is fulfilled for on-site
  energies of the order of $10^3$ eV.}.

\begin{figure}[htbp]
        \centering
        \includegraphics[width=0.45\columnwidth]{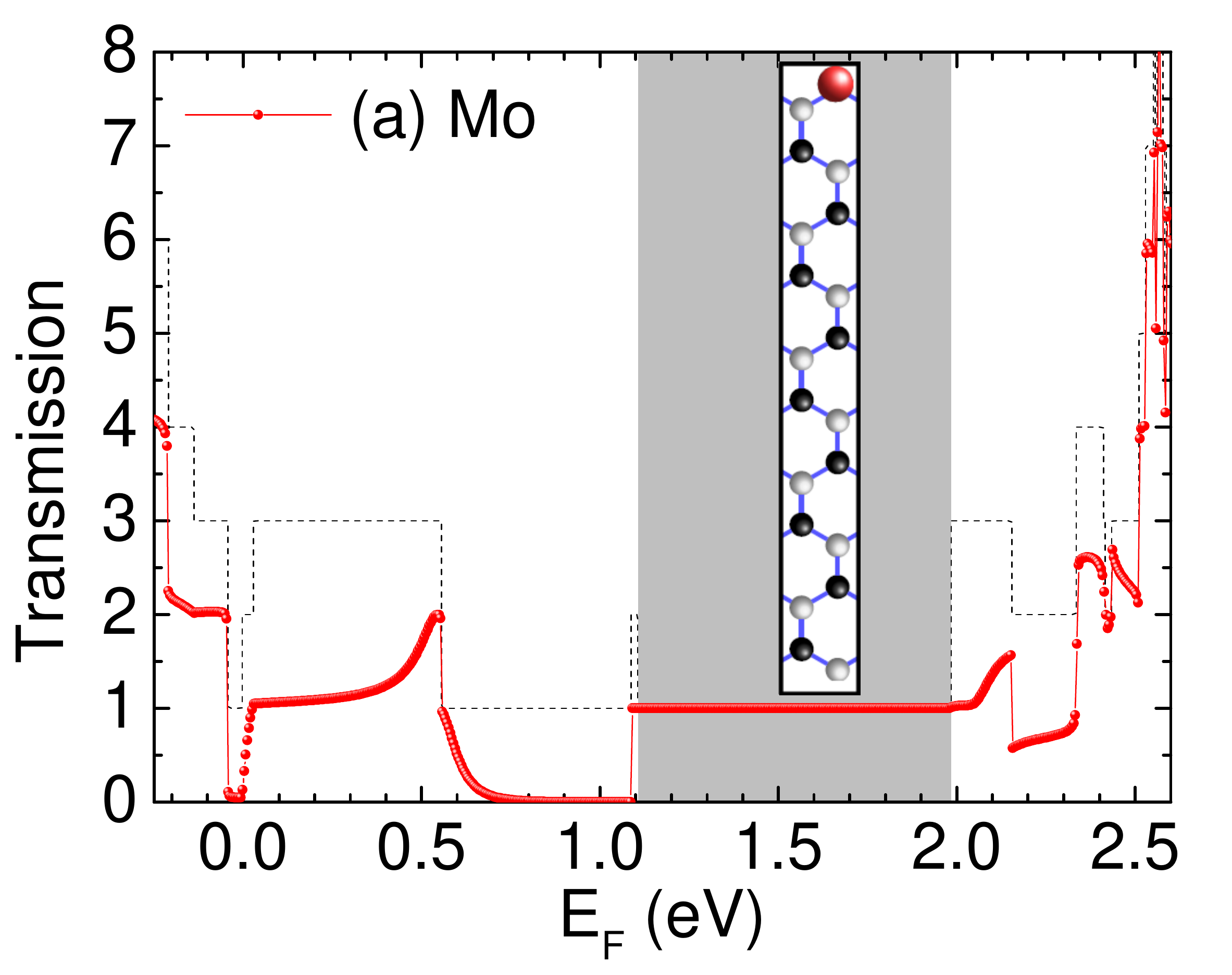} 
				\includegraphics[width=0.45\columnwidth]{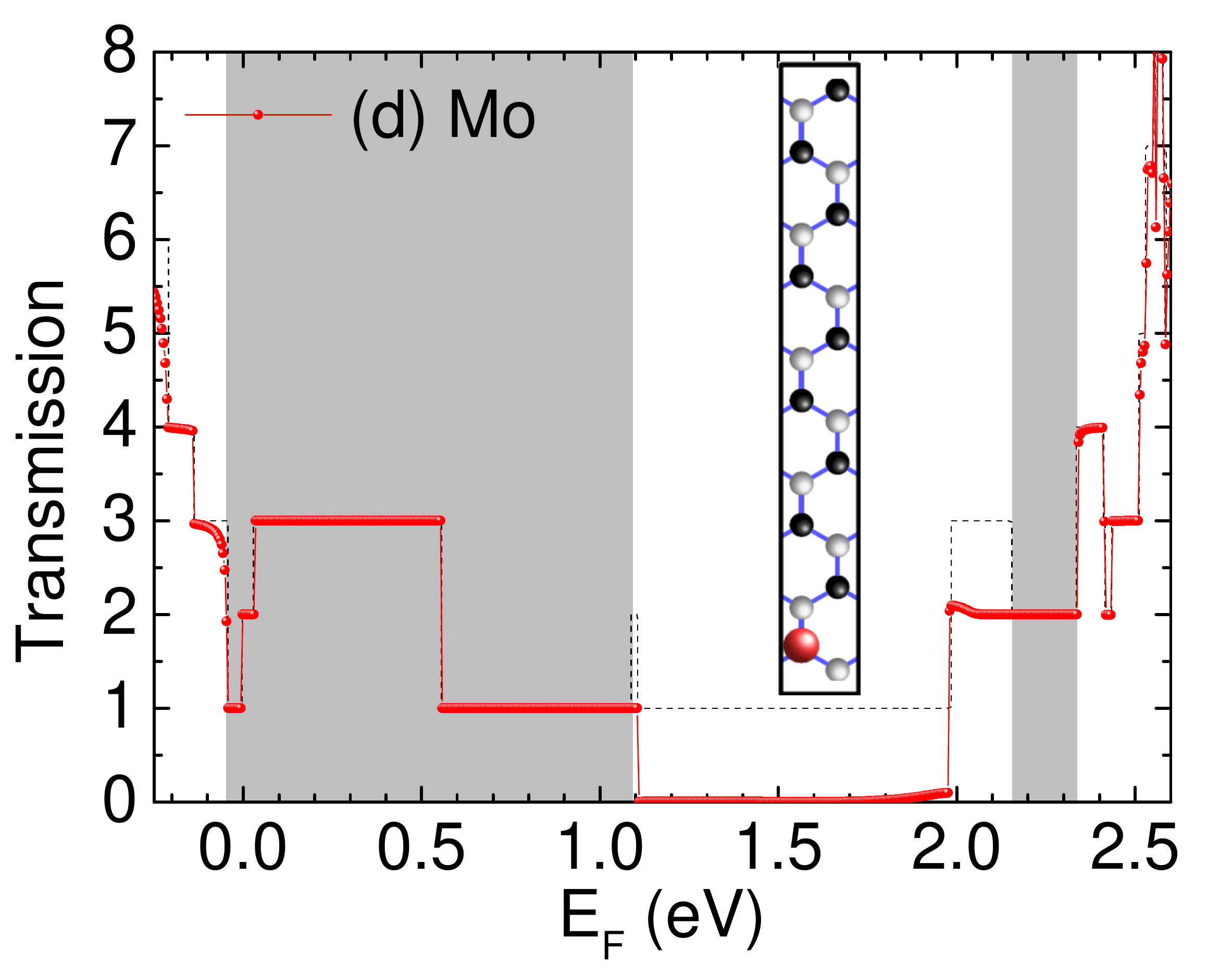}
				\includegraphics[width=0.45\columnwidth]{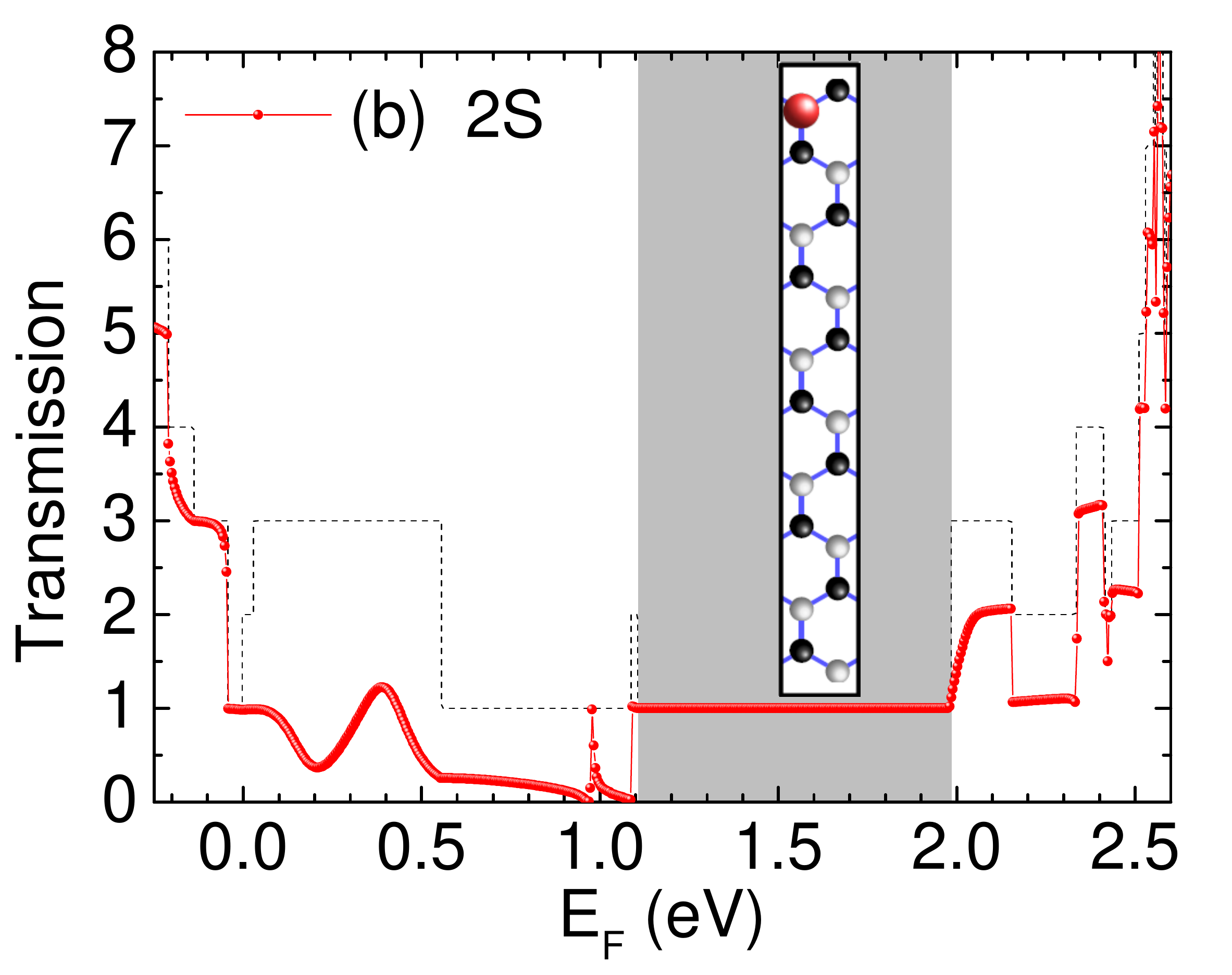} 
				\includegraphics[width=0.45\columnwidth]{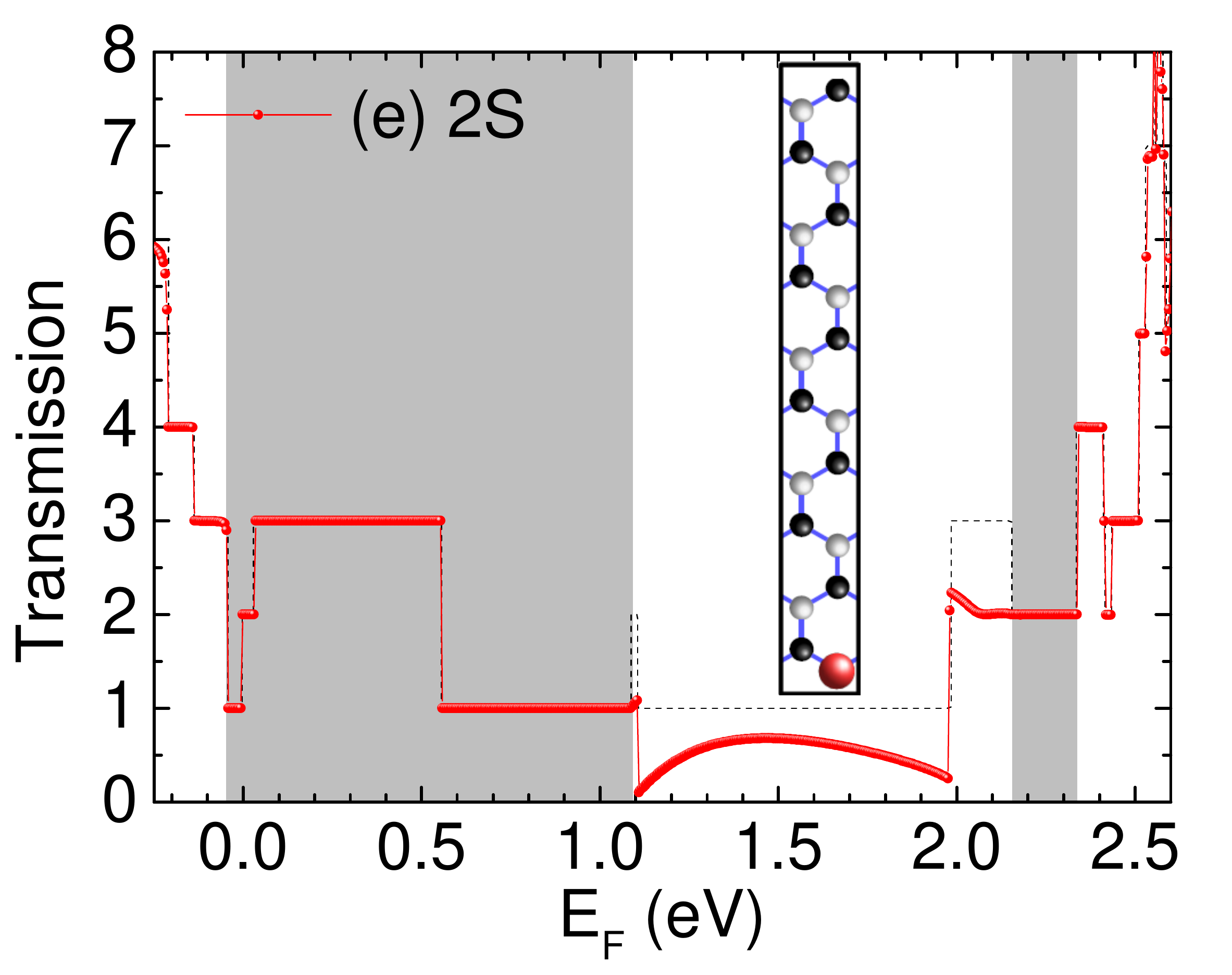}
				\includegraphics[width=0.45\columnwidth]{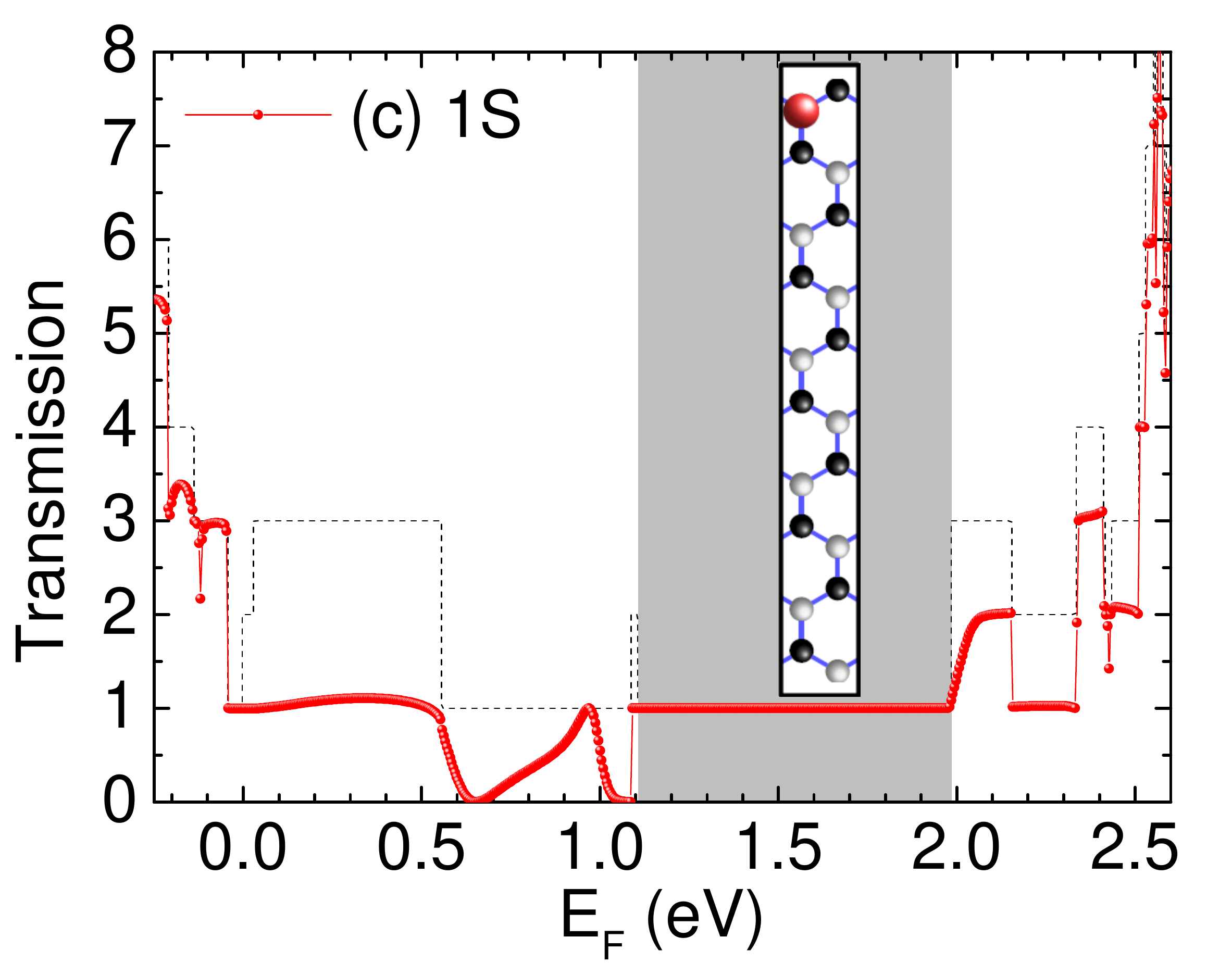} 
				\includegraphics[width=0.45\columnwidth]{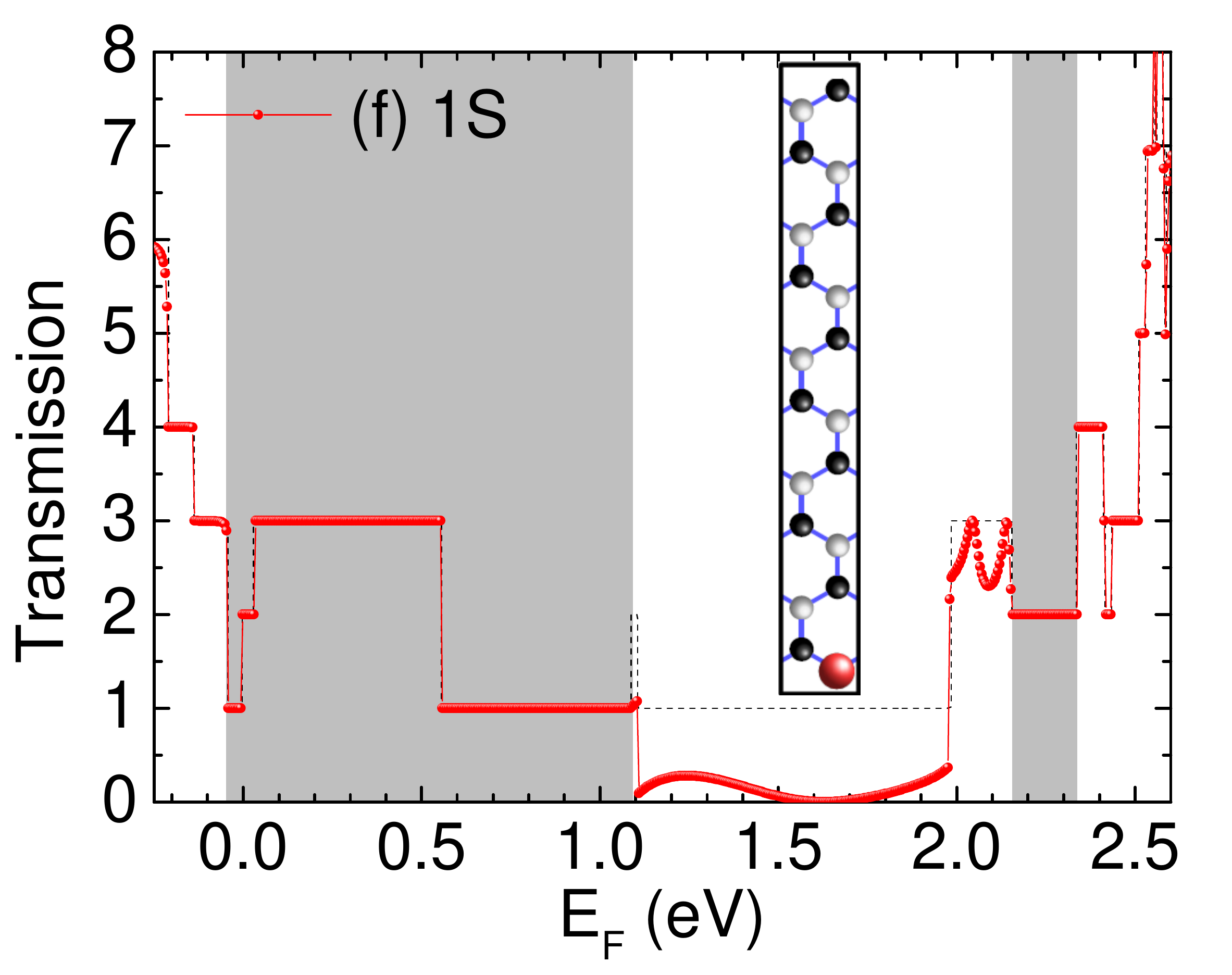}
 \caption{(Color online) Electronic transmission in a 10-Z-MoS$_2$-NR
   as a function of the Fermi energy for the clean case (black curve)
   and in the presence of a vacancy at the edge (red curve). The
   nanoribbon contains $M=100$ cells and the vacancy is present in the
   middle cell that is equally distant to the contacts. The ribbon
   contains $20$ atoms in each cell. The insets indicate the vacancy
   position. The panels on the left show results for vacancies in the
   top edge (Mo-terminated edge): (a) one Mo vacancy, (b) two S
   vacancies, one in the bottom plane and one in the top plane, and (c)
   one S vacancy in the upper or bottom plane. The panels on the right
   show results for vacancies in the bottom edge (S-terminated edge):
   (d) one Mo vacancy, (e) two S vacancies, one in the bottom plane and
   one in the top plane, and (f) one S vacancy in the upper or bottom
   plane. The gray areas mark the perfect transmission regions due to
   the states at the opposite edge.}
 \label{fig:one_vacancy}
\end{figure}

We perform a systematic study of the effect of a single vacancy on
$T_{RL}$ covering all possible vacancy configurations in the
nanoribbon unit cell (transverse slice). We consider geometries where
the vacancy is equidistant from source and drain contacts to avoid
effects due to the coupling to the leads. Here we consider ribbons of
width N=$10$. We calculated the conductance for much wider ribbons
(N=$50$) for a few cases and obtained very similar results in the gap
region, ruling out finite-size effects.

We describe the main findings in Fig.~\ref{fig:one_vacancy}. The
figure illustrates the effect only for vacancies placed at the nearest
edge atomic positions, since we find that the reduction in the
transmission becomes progressively smaller as the vacancy is moved
away from the nanoribbon edges. We show the transmission for six
different vacancy configurations, namely, single S vacancies (top and
bottom vacancies are equivalent due to symmetry), double S vacancies
(top and bottom planes), and single Mo vacancies near each edge. We
recall that $0 \alt E \alt 2.25$ eV corresponds to the bulk gap. The
gray areas mark the energy intervals of perfect transmission. The
corresponding edge states are robust against the presence of
vacancies. This can be understood as follows: a vacancy on the
Mo-terminated edge, as shown in Figs.~\ref{fig:one_vacancy}(a),
\ref{fig:one_vacancy}(b), and \ref{fig:one_vacancy}(c), suppresses the
transmission of the bands $0$, $2$, and $3$ while band $1$ remains
unaffected with perfect transmission. On the other hand, vacancies on
the S-terminated edge strongly suppress the transmission of band $1$
and do not affect bands $0$, $2$, and $3$, as seen in
Figs.~\ref{fig:one_vacancy}(d), \ref{fig:one_vacancy}(e), and
\ref{fig:one_vacancy}(f). These results indicate that the propagation
of band $1$ occurs mainly through the S-terminated edge, while bands
$0$, $2$ and $3$ propagate through the Mo-terminated edge. Note that
for $2 \alt E \alt 2.15$ eV both bands $0$ and $1$ are accessible [see
Fig.~\ref{fig:LDOS_clean}(c)] and hence any vacancy configuration
near the nanoribbon edges strongly suppresses the transmission.
Curiously, Fig.~\ref{fig:one_vacancy} indicates that S di vacancies have a smaller effect on the conductance than S mono vacancies.

\begin{figure}[htbp]
        \centering
        \includegraphics[width=0.9\columnwidth]{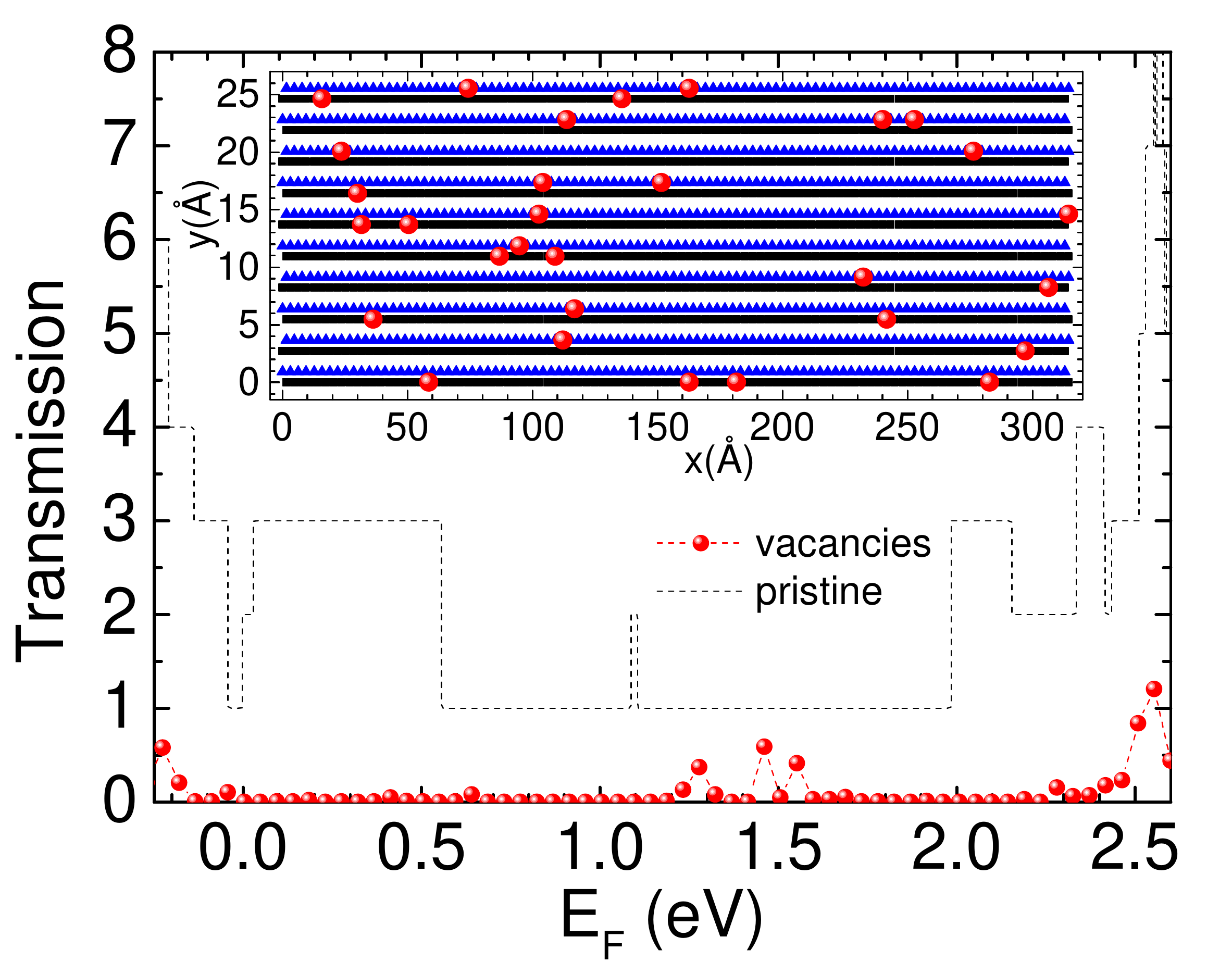}
 \caption{(Color online) Electronic transmission in a 10-Z-MoS$_2$-NR
   as a function of the Fermi energy in the presence of random
   vacancies. The dashed black lines indicate perfect transmission of
   the pristine ribbon. The nanoribbon contains $M=100$ cells, with
   $30$ atoms in each cell. The inset indicates the vacancy
   positions (red balls) distributed in both S (black squares) and Mo
   (blue triangles) lattices.}
 \label{fig:transmission_vacancies}
\end{figure}

In summary, we find that the electronic transport results in the
presence of a single vacancy are consistent with the LDOS of the
pristine 10-Z-MoS$_2$-NR presented in Fig.~\ref{fig:LDOS_clean}.
Vacancies near the nanoribbon edges dramatically suppress the
conductance. Thus, short-range disorder breaks the valley topological
protection of the metallic states in zigzag MoS$_2$ nanoribbons.
Short-range
defects, such as vacancies, enable intervalley scattering processes
and hence, backscattering. This reasoning supports the picture that
the system behaves like a trivial metal. Are the observations
consistent with the robustness of edge states recently experimentally
reported in Refs.~\cite{Zhang2014} and \cite{Koos2016}? This is the question
we address next.

Let us now analyze the electronic transport in a more realistic case
by placing random vacancies in the central region.
Figure~\ref{fig:transmission_vacancies} shows the electronic
transmission of a disorder realization for a total vacancy
concentration (Mo and double S) of $0.15\%$ for 10-Z-MoS$_2$-NR. As
expected, a finite vacancy concentration is extremely detrimental to
the electronic transport in the bulk gap energy region. This is a
manifestation of Anderson localization in a one-dimensional disordered
metal.

\begin{figure}[htbp]
        \centering
        \includegraphics[width=0.99\columnwidth]{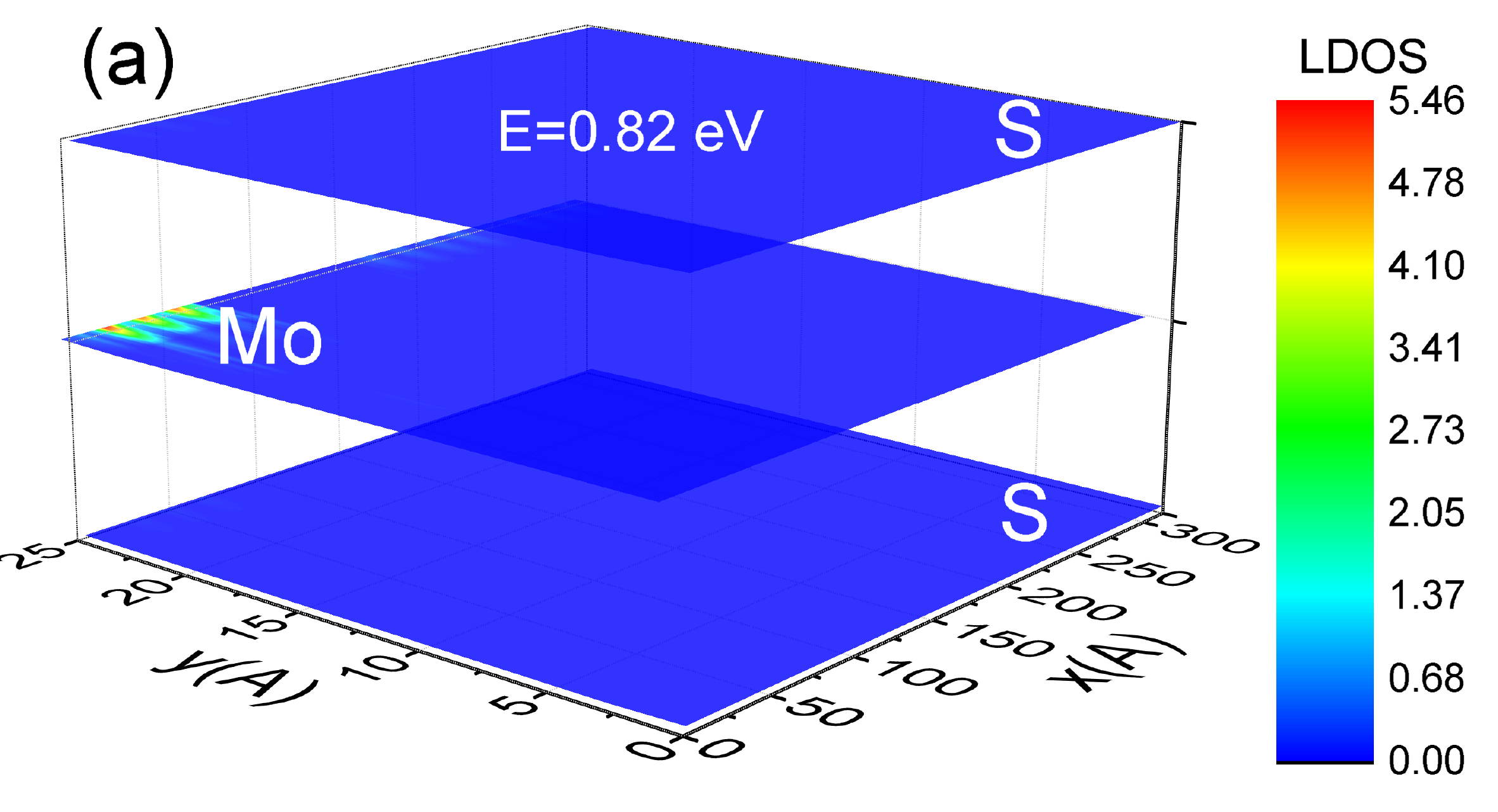}
				\includegraphics[width=0.99\columnwidth]{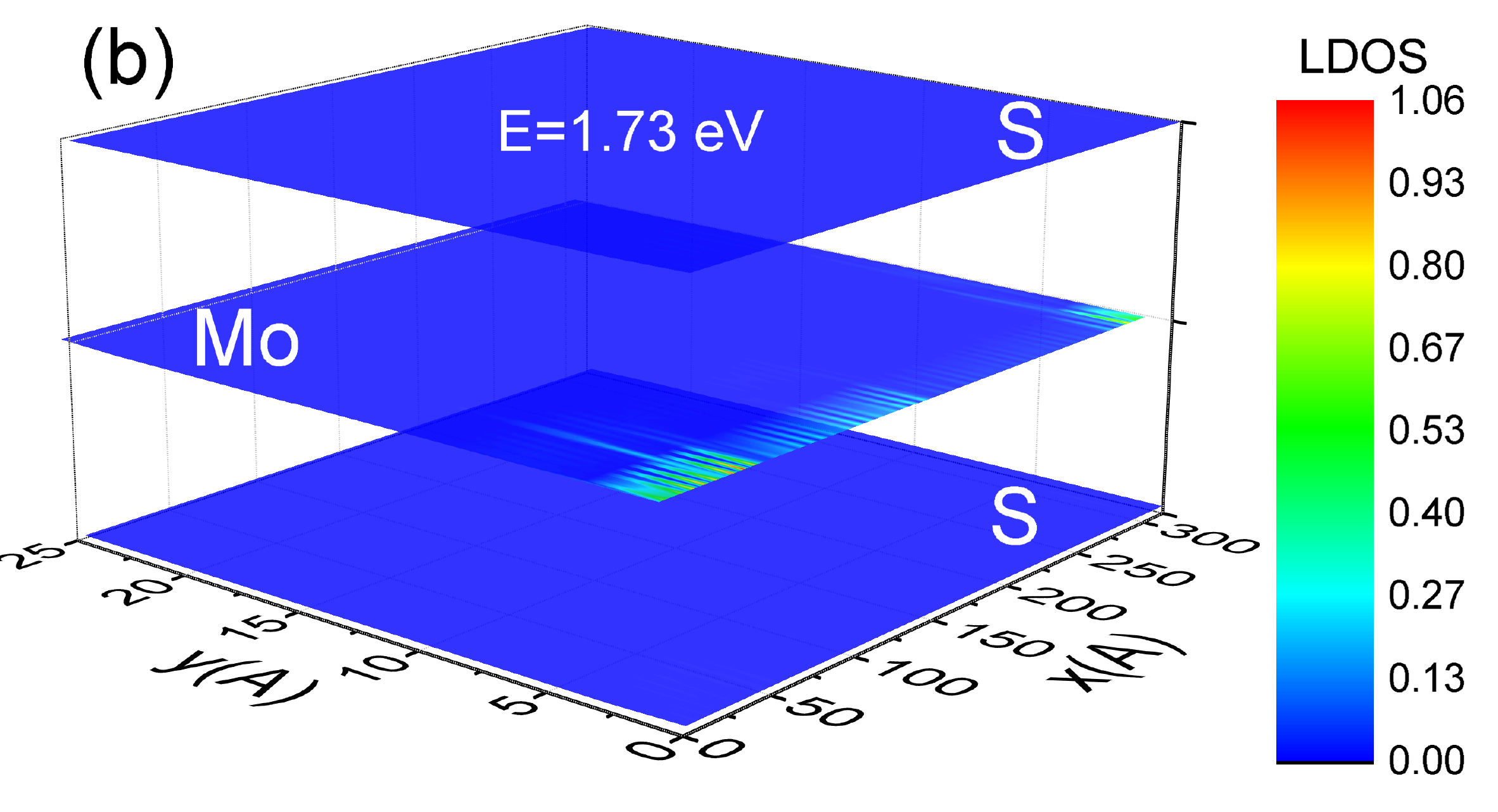}
 \caption{(Color online) LDOS at the energies (a) $E=0.82$ eV and (b)
   $E=1.73$ eV for the 10-Z-MoS$_2$-NR with added random vacancies, as
   shown in Fig.~\ref{fig:transmission_vacancies}. The ribbon edge at
   $y=0$ is S terminated while the opposite one is Mo terminated. The
   image in each plane corresponds to a single atomic nature, S or Mo,
   as indicated by the labels in each plane.  }
 \label{fig:LDOS_vacancies}
\end{figure}

Figure \ref{fig:LDOS_vacancies} shows the LDOS for the same disorder
configuration mentioned above (see
Fig.~\ref{fig:transmission_vacancies}).
We analyze two selected energies, namely, $E=0.82$ eV
[Fig.~\ref{fig:LDOS_vacancies}(a)] and $E=1.73$ eV
[Fig.~\ref{fig:LDOS_vacancies}(b)], that correspond to states
concentrated at opposite edges. The vacancies near the edges
create depletion regions in LDOS, localizing the wave functions and
hindering the transport. Surprisingly, we find that the vacancies also
induce the appearance of regions with high LDOS at the edges that
exceed the maximum LDOS of the pristine system in
Fig.~\ref{fig:LDOS_clean}. We find similar trends for other energies,
justifying the wide transport gap ($\approx 2.25$ eV) found in
Fig.~\ref{fig:transmission_vacancies}.

\subsubsection{Long-range scattering}

Let us now consider the case of long-range impurity disorder, which
can be caused, for instance, by an inhomogeneous charge distribution
in the substrate \cite{Lima16} or by disordered
ripples \cite{Touski2016}. To study the degree of protection of the
edge states against long-range disorder, we place a Gaussian on-site
potential centered at $\mathbf r_{0}$, namely,
\begin{align}
	V(\mathbf r) = V_0 e^{-|\mathbf r - \mathbf r_0|/2d^2},
\end{align}
where $V_0$ and $d$ are the impurity strength and range, respectively.
Here ${\mathbf r}_{0}$ is taken at one of the system edges and placed
equidistantly to the contacts. By varying $d$ we can access different
scattering range regimes and determine the corresponding effect on the
electronic transmission of the system. In graphene deposited on
SiO$_{2}$, the typical rms of $V(\mathbf r)$ is $100$ meV \cite{Xue11}.
In the absence of a better experimental guidance in the case of
MoS$_2$, we take $V_0=0.1$ eV.

\begin{figure}[htbp]
        \centering
        \includegraphics[width=0.90\columnwidth]{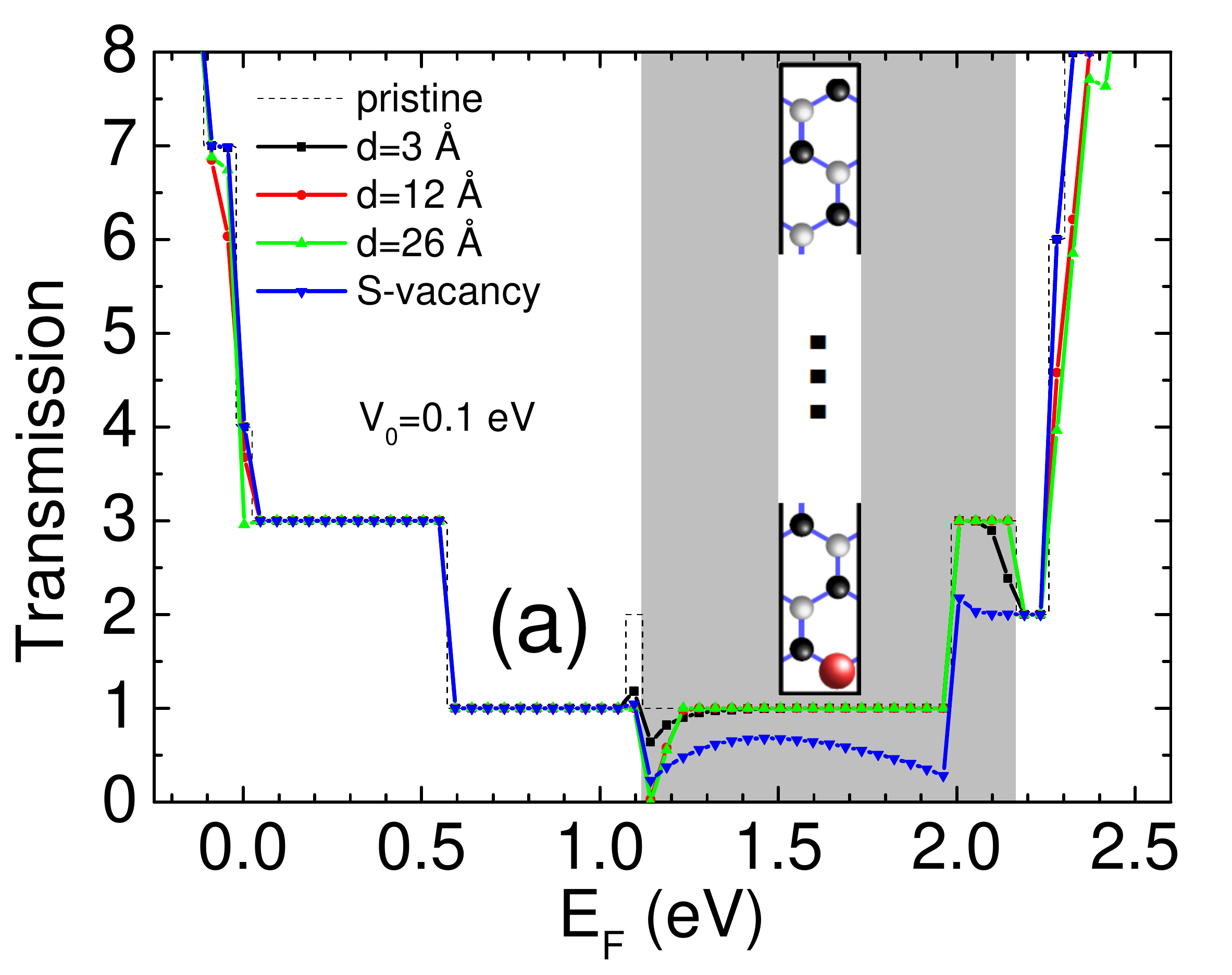} 
				\includegraphics[width=0.90\columnwidth]{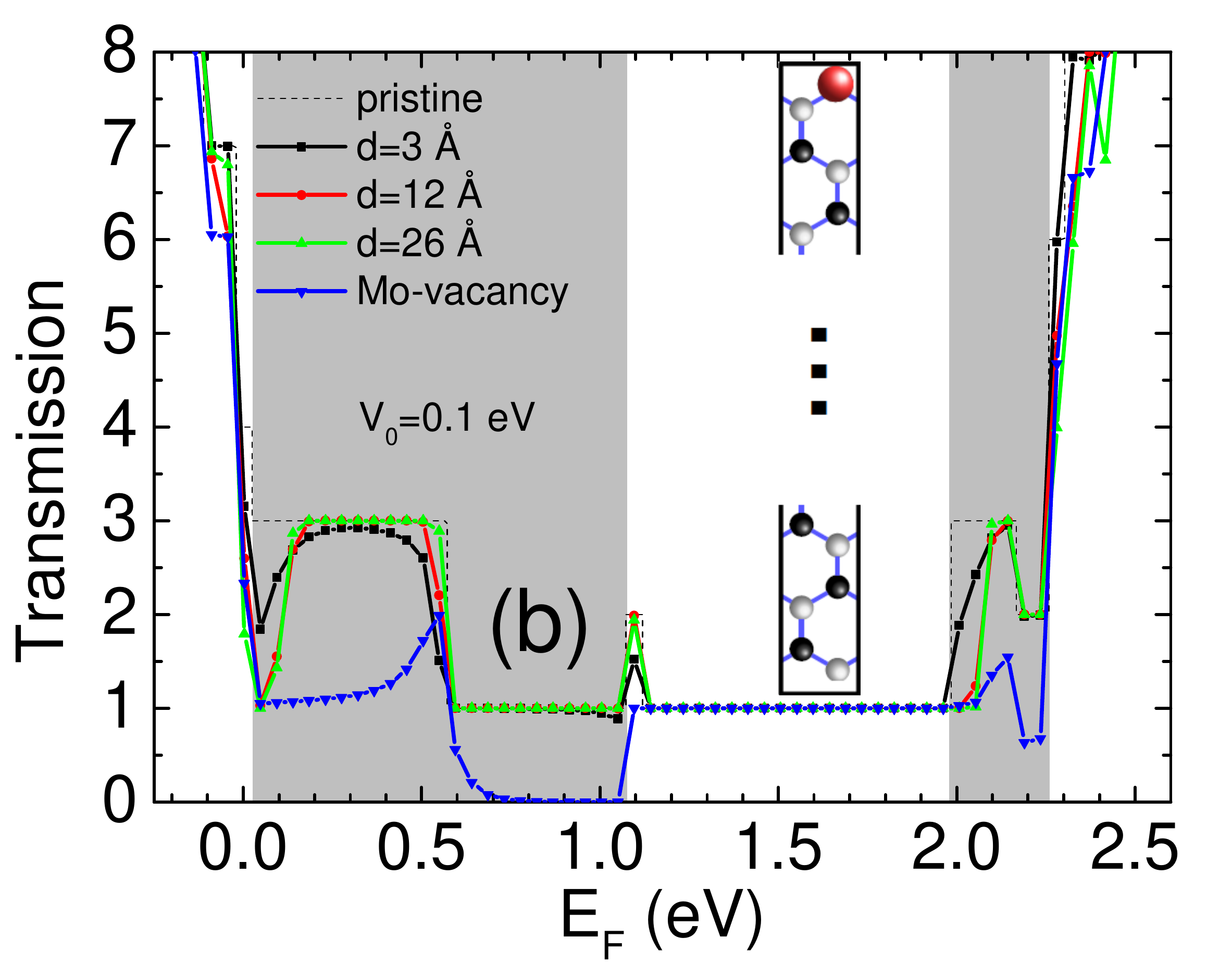}
 \caption{(Color online) Electronic transmission in a 30-Z-MoS$_2$-NR
   as a function of the Fermi energy. We compare the transmissions due
   to vacancies and the on-site impurity for ranges $d=3$\,\AA,
   $12$\,\AA, and $26$\,\AA. The nanoribbon contains $N=100$ cells, and
   the vacancy is present in the middle cell that is equally distant
   from the contacts. The ribbon contains $20$ atoms in each cell.
   The insets indicate the vacancy and impurity position. In (a) we
   show the results for defects in the top edge at the S atom, while
   in (b) the impurity is on the Mo atom. The gray areas mark the
   energy regions where the states are characterized by a propagation
   at the edge where the defect is placed.}
 \label{fig:longrange}
\end{figure}

Let us consider the case of an impurity placed at the S-terminated
edge. Figure \ref{fig:longrange}(a) shows the corresponding
transmission for three impurity configurations, namely, $d=3$ \AA,
$d=12$ \AA, and $d=26$ \AA\, for $V_0=0.1$ eV in a 30-Z-MoS$_2$-NR.
We find a nearly perfect transmission quantization for all long-range
impurity configurations. We find the same behavior for the impurity at
the Mo-terminated edge. In contrast, the shaded areas in
Fig.~\ref{fig:longrange}(b) indicate the energy windows where the
transmission is not perfect, corresponding to edge states localized at
the same edge as the impurity center.

There are also energy windows where we find a stronger suppression of
the conductance. For instance, for energies around $E_F=1.05$ eV,
backscattering processes require a small momentum transfer [see Fig.~\ref{fig:bands}(b)]. As a consequence, long-range impurity enables
a transmission suppression as shown in Fig.~\ref{fig:longrange}. A
similar behavior can be seen at $0\alt E\alt 0.5$ eV, where the
band structure allows for several small momentum backscattering
processes.

\section{Conclusions}
\label{sec:conclusions}

In this paper we use the tight-binding model recently proposed by our
group \cite{Ridolfi2015} to systematically study the electronic
structure and transport properties of monolayer MoS$_2$
nanoribbons. Our results reproduce qualitatively the DFT band-structure calculations when limited to narrow and pristine
nanoribbons. We argue that it is necessary to consider the full
tight-binding Hamiltonian, with even and odd parities, for an accurate
description of nanoribbon electronic states within the bulk gap energy
window. By doing so, we observe the appearance of an odd-parity band
close to the Fermi level at charge neutrality for both kinds of edge
terminations. 
Such bands contribute significantly to the conductance.
We analyze the edge nature of the states inside the bulk gap in the
zigzag case through the orbital composition of the eigenstates at a
given energy and $k$ point. Interestingly, we find that the metallic
bands correspond to states localized at a single edge independent of the nanoribbon width and disorder, as we
explicitly show in our LDOS calculations.

We study the effect of short- and long-range disorder on the
conductance and LDOS of zigzag nanoribbons. We find that even a modest
concentration of vacancies close to the edges can cause a large
transmission suppression, particularly within the bulk gap energy
window. In contrast, weak long-range scattering does not have a
significant effect on the conductance. We interpret these results in
terms of the nature of the metallic bands and their protection against
disorder, since the system is robust against intra valley scattering
processes.
Despite the differences in the calculated band structures, our findings are consistent with the classification of the weak topological insulator proposed in Ref.~\cite{Rostami15}. 

In summary, our analysis indicates that short-range defects such as
vacancies and edge roughness dramatically suppress the conductance of
MoS$_2$ nanoribbons, as they create regions at the edge with
negligible LDOS. Surprisingly, there are also regions with enhanced
LDOS due to the metallic edge states that are quite robust against
disorder. This is consistent with recent scanning tunneling microscopy and spectroscopy experiments that
observed a significant localized LDOS at the (rough) edge of a MoS$_2$
sample \cite{Zhang2014,Koos2016}. To verify that the LDOS is not
continuous at the edge and it is not due to adsorbates or molecules,
more experimental input is needed.

\acknowledgements This work was supported by the Brazilian funding
agencies CNPq, CAPES, and FAPERJ. E.M. thanks the Instituto de
F\a'{i}sica at Universidade Federal Fluminense for their hospitality.

\bibliography{TMDs}

\end{document}